\documentclass[aps,amsmath, amssymb, prd, twocolumn, superscriptaddress,floatfix,showpacs,10pt]{revtex4-1}
\usepackage{graphicx}  
\graphicspath{{figures/}} 
\usepackage{natbib}

\usepackage{hyperref}			

\usepackage{sistyle}
\usepackage{enumitem}          
\usepackage{color}
\usepackage{cancel}
\usepackage{stmaryrd}

\usepackage{mathtools}
\newcommand{\eqcolon}{\mathrel{\resizebox{\widthof{$\mathord{=}$}}{\height}{ $\!\!=\!\!\resizebox{1.2\width}{0.8\height}{\raisebox{0.23ex}{$\mathop{:}$}}\!\!$ }}}
\newcommand{\coloneq}{\mathrel{\resizebox{\widthof{$\mathord{=}$}}{\height}{ $\!\!\resizebox{1.2\width}{0.8\height}{\raisebox{0.23ex}{$\mathop{:}$}}\!\!=\!\!$ }}}

\begin{document}

\title{Pulse redshift of pulsar timing array signals for all possible gravitational wave polarizations in modified general relativity}

\author{Adrian \surname{Bo\^itier}}
\email[]{boitier@physik.uzh.ch}
\affiliation{Physik-Institut, Universit\"at Z\"urich, Winterthurerstrasse 190, 8057 Z\"urich}

\author{Shubhanshu \surname{Tiwari}}
\email[]{stiwari@physik.uzh.ch}
\affiliation{Physik-Institut, Universit\"at Z\"urich, Winterthurerstrasse 190, 8057 Z\"urich}

\author{Lionel \surname{Philippoz}}
\email[]{plionel@physik.uzh.ch}
\affiliation{Physik-Institut, Universit\"at Z\"urich, Winterthurerstrasse 190, 8057 Z\"urich}

\author{Philippe \surname{Jetzer}}
\email[]{jetzer@physik.uzh.ch}
\affiliation{Physik-Institut, Universit\"at Z\"urich, Winterthurerstrasse 190, 8057 Z\"urich}

\date{\today}

\begin{abstract}
Pulsar timing arrays (PTA) have the promise to detect gravitational waves (GWs) from sources which are in a unique frequency range of $10^{-9}$- $10^{-6}$ Hz. This in turn also provides an opportunity to test the theory of general relativity in the low frequency regime. The central concept of the detection of GWs with PTA lies in measuring the time of arrival difference of the pulsar signal due to the passing of GWs i.e. the pulses get red-shifted. In this paper we provide a complete derivation of the redshift computation for all six possible polarizations of GW which arise due to the modifications to general relativity.  We discuss the smoothness of the redshift and related properties at the critical point, where the GW source lies directly behind the pulsar. From our mathematical discussion we conclude that the redshift has to be split differently into polarization part (pattern functions) and interference part, to avoid discontinuities and singularities in the pattern functions. This choice of pattern functions agrees with the formula one uses for interferometers with a single detector arm.
Finally, we provide a general expression which can in principle be used for pulsars and GW of any frequency without invoking the low frequency assumption and using said assumption we develop the expression up to first order in the strain and find correction terms to the canonical redshift formula.
\end{abstract}


\maketitle

\section{Introduction}
The discovery of GW with LIGO and Virgo detectors of compact objects has provided  unprecedented understanding of not only these compact objects but also broadly about general relativity (GR) and cosmology \cite{GWTC-1, GWTC-TGR}. These instruments are sensitive to high frequency GW i.e. $10-10^4$ Hz, and the sources which can be probed are stellar mass black hole and neutron star binaries just before merging, supernovae etc. Also there are space based experiments like LISA which will probe the medium frequency range $10^{-4} - 10^{-1}$ Hz, where the merger of supermassive mass black hole binaries will be detected \cite{LISA}. PTA probes much lowerer frequencies  $10^{-9}$- $10^{-6}$ Hz, the expected source here is supermassive black holes in the early stage of inspirals. Hence PTA provides a complementary view of the gravitational waves sky and provides the opportunity to test GR in nanohertz regime \cite{PTA-TGR}.  \\ 

Pulsar timing array experiments \cite{NANOGRav11yr,IPTA, EPTA, PPTA} use the stability of the pulse periods of neutron stars which happen to point their jets in our direction. This causes a lighthouse effect, that can be used as a very precise clock. If a gravitational wave from a distant supermassive black hole binary or a gravitational wave background passes through our galaxy, the pulse gets red shifted due to the change in the distance between Earth and the pulsar and thus a change in the travel time of the photons. The deviation from a regular pulse would allow us to detect a gravitational wave.\\

The pattern functions which describe the sensitivity of the Earth-pulsar system as a function of sky directions, used for PTA’s revealed discontinuities for $+$ and $\times$ polarization and even poles in the case of $x, y$ and $l$, as presented in \cite{Sesana} for example. The responsible term in these expressions is suspiciously independent of polarization. This definition of the pattern functions also seems to be not harmonious with the definition of pattern functions normally used for interferometers. This motivated us to look more closely into the derivation of the redshift formula since the resulting redshift is smooth.\\ 

In this work we provide a a complete and independent derivation for all six polarizations for the pulse redshift in section~\ref{sec: PulseRedshift}. This enables us to develop the expression for the pulse redshift up to first order in the GW frequency for a fixed pulsar period. Till now only the zeroth order expression was available from \cite{Detweiler}. From our methodology one could calculate higher order terms. \\


In his publication on pulsar timing measurements \cite{Detweiler} Detweiler presents the results of a slight variation of the derivation by Estabrook and Wahlquist \cite{Est&Wahl}. Estabrook and Wahlquist derived the contribution of GW to the Doppler shift of a continuously emitted and coherently retransmitted sinusoidal electromagnetic signal, which is used to track distant spacecraft. The expressions for the “one-way redshift”, which they use and correctly describes the PTA redshift have been derived by W. J. Kaufmann who compared the energies at emission and reception of a photon to calculate the photon redshift, in his publication \cite{Kaufmann} which is motivated by the comparison of a perturbed flat spacetime geometry to a refracting medium.\\

Maggiore \cite{Maggiore2} calculates the pulse redshift via the arrival time difference. However, it is assumed there, that photons would in the chosen reference frame travel on straight lines. Our solution~\eqref{eq.: pGeod} to the linearized geodesic equations without prior assumptions show, that this is an oversimplification.\\

We start in section~\ref{sec: Reasonablility} by pointing out certain subtleties concerning  the commonly used pattern functions. We point out, that there are poles for all polarizations that have a longitudinal component and that the positions of blind spots and maxima do not agree with intuition in~\ref{subsec: FA's}. Then we turn our attention to the symmetries in the gravitational waves coming from their helicity in~\ref{subsec: H&D} and prove that since the signal has these symmetries they also have to be apparent in the response of the detectors measuring said signal.\\

Since the issue of a maxima of a pattern function lying in a blind spot is argued away with negative interference in \cite{Sesana}, we dedicate section~\ref{sec: DetectorTensor} to explain, why these two effects have nothing to do with each other and can therefore not be used to cure the problem.\\

We finally give our own derivation in section~\ref{sec: PulseRedshift} and a discussion in section~\ref{sec. discussions}.

\section{Antenna pattern function : A closer look}\label{sec: Reasonablility}

\subsection{Poles and jumps}\label{subsec: FA's} 
A general formula for the antenna pattern functions for $+$ and $\times$ polarization is given in \cite{Sesana} and \cite{Maggiore2}:
\begin{equation}\label{eq: FalseFA}
F^A(\hat{\Omega}) = \frac{1}{2}\frac{\hat{p}^i\hat{p}^j}{1+\hat{p}^i\hat{\Omega}_i}e^A_{ij}(\hat{\Omega}),
\end{equation}
where $\hat{p}$ is the unit direction vector in which the pulsar lies, $\hat{\Omega}$ is the direction in which the GW travels and $A \in \lbrace +,\times\rbrace$.\\

It is immediate to generalize this to additional polarizations, coming from modified GR, by inserting the polarization tensors of these polarizations.\\
If we have a longitudinal part in the polarization, as it is the case for vector-$x$, $y$ and of course the pure longitudinal polarization $l$, then this formula leads to poles. If we use the same choice of reference frame and convention for $\theta$ as in \cite{Sesana} (GW incident in direction $\hat{\Omega} = (0,0,-1)$ and pulsar located in direction $\hat{p} = (\sin\theta,0,\cos\theta)$) in the example of $x$- and $l$-polarization, we get:
\begin{align}
F^x(\theta) = \frac{\sin\theta\cos\theta}{1+\cos\theta},	&&	F^l(\theta) = \frac{1}{2}\frac{\cos^2\theta}{1+\cos\theta}.
\end{align}

Both expressions have a pole at $\theta = \pi$, which we show by expanding around the pole: $\theta = \pi + \delta\theta$;
\begin{align}
\lim_{\theta\rightarrow\pi}\frac{\sin\theta\cos\theta}{1+\cos\theta} &= \lim_{\delta\theta\rightarrow 0}\frac{\delta\theta(-1+\frac{1}{2}\delta\theta^2)}{1-1+\frac{1}{2}\delta\theta^2} \rightarrow -\infty, \notag \\
\lim_{\theta\rightarrow\pi}\frac{1}{2}\frac{\cos^2\theta}{1+\cos\theta} &= \frac{1}{2}\lim_{\delta\theta\rightarrow 0} \frac{1-\delta\theta^2+\frac{1}{4}\delta\theta^4}{\frac{1}{2}\delta\theta^2} \rightarrow \infty.
\end{align}

\begin{figure}[h!]
	\centering\includegraphics[width=\linewidth]{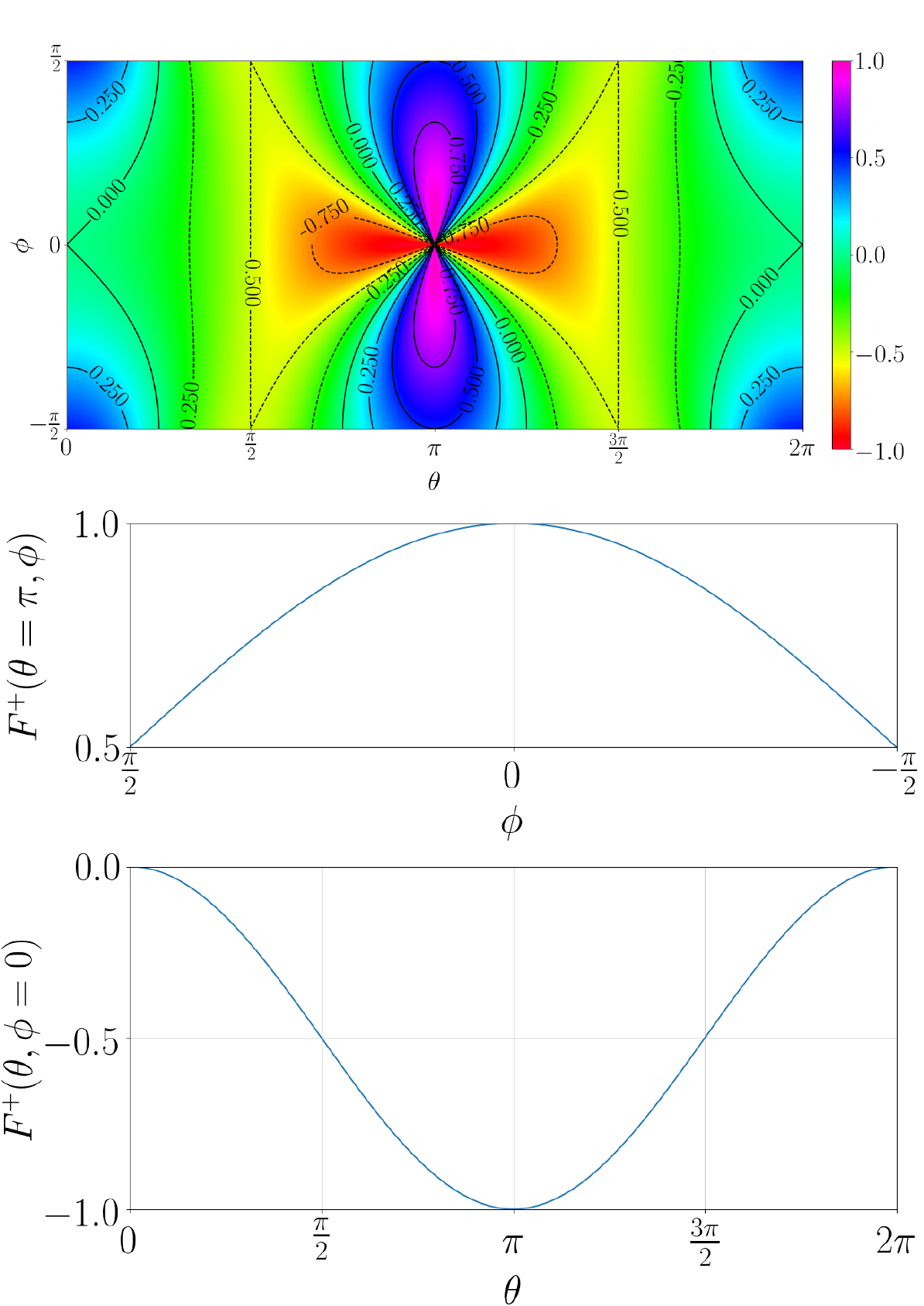}
\caption{The angular pattern function for the $+$ polarization $F^+(\hat{\Omega})$ with its
	discontinuity in the middle of the image and its plot along the $\phi$- and $\theta$-axis 
	through the discontinuity.}
\label{fig: F+(Omega)}
\end{figure}

The pattern functions for the $+$ and $\times$ polarization do not have any poles, but do not reflect the expected response of a pulsar detector either. Since these two are pure transverse polarizations we should have a blind spot when the wave travels in the direction of the pulsar or the opposite i.e. $\theta = 0$ or $\pi$. Additionally we would expect a maximum at $\theta = \pm \frac{\pi}{2}$ in the case of the $+$ polarization.\\
We do have a blind spot at $\theta = 0$, but we find a maximum (discontinuity) at $\theta = \pi$. We constrain ourselves to the plane in which the Earth-pulsar line and one axes of the $+$ polarization lies and demonstrate the result in this case:
\begin{equation}
F^+(\pi) = \lim_{\theta\rightarrow\pi} \frac{1}{2}\frac{\sin^2\theta}{1+\cos\theta} = \frac{1}{2}\lim_{\delta\theta\rightarrow 0} \frac{\delta\theta^2}{\frac{1}{2}\delta\theta^2} = 1.
\end{equation}
For the $\times$ polarization we would have to tilt the plane by $\frac{\pi}{4}$, to get the same result.\\
In fact, we get different limits, when we approach $\theta = \pi$ from different directions. This means that the response function is not continuous at that point (Fig.~\ref{fig: F+(Omega)}).

The discontinuity at the poles come from the degeneracy of the polar coordinates at these points. The one in the middle of the plot does not arises by the choice of coordinates as the polar coordinates are well defined and smooth at the equator. Therefore we conclude, that~\eqref{eq: FalseFA} is not the optimal quantity to be considered to describe the response of a detector towards polarization. Also more practically it causes numerical errors whilst computing, since cancelling poles with zeroes is unstable.\\

\subsection{Helicity of GWs}\label{subsec: H&D}
Gravitational waves have a helicity of 2. Geometrically this means that if one rotates the wave by $\frac{2\pi}{2}$ we always get the same wave. The consequence of this helicity is that all polarizations reshape a circle of test masses into an ellipse (the spatial case of a circle for the breathing polarization). An ellipse has a mirror symmetry with respect to its two axes. These symmetries should be reflected in the response functions.\\

The Hellings \& Downs curve $\alpha_{ij}$ \cite{H&D} (see also Maggiore \cite{Maggiore2}, p. 732, eq.(23.39)) describes the overlap reduction function of the tensor mode under the short wavelength approximation. It is the direction averaged geometric part of the cross correlation of two signals of different detectors (pulsars) which we denote as $i$ and $j$:
\begin{align}
&\alpha_{ij}(\gamma_{ij}) = \frac{1}{4\pi}\int_{\mathcal{S}^2} F^+_i(\hat{\Omega})F^+_j(\hat{\Omega}) + F^\times_i(\hat{\Omega})F^\times_j(\hat{\Omega}) d\hat{\Omega} \notag \\
&= \frac{1-\cos\gamma_{ij}}{2}\ln\left( \frac{1-\cos\gamma_{ij}}{2} \right) - \frac{1}{6}\frac{1-\cos\gamma_{ij}}{2} + \frac{1}{3},
\end{align}
where $\gamma_{ij}$ is the angle between the two pulsars and $F^A$ are given by eq.~\ref{eq: FalseFA}.\\

\begin{figure}
	\centering\includegraphics[width=\linewidth]{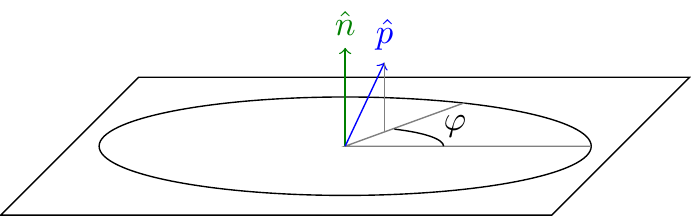}
	\caption{Sketch of the plane in which the gravitational wave distorts space with surface normal vector $\hat{n}$ and the projection of the detector arm $\hat{p}$ onto it.}
	\label{fig: DistortionPlane}
\end{figure}

To exploit the symmetries, we decompose the pattern functions in a different way.\\
Only the projection of the detector arm $P\hat{p}$ into the plane in which the ellipse lies~\ref{fig: DistortionPlane} contributes to the signal. The orthogonal part stays invariant under the influence of the GW. The part in the plane is described by a function $f^A$ that maps the projected angle to the deviation from the circle at that angle.
\begin{equation}
	F^A(\hat{\Omega}) = \sqrt{1-(\hat{p}\cdot\hat{n})^2}f^A(\varphi),
\end{equation}
where $\hat{n}$ is the surface normal of the ellipse plane and $\varphi$ the angle between a semi-major axis and $P\hat{p}$.\\

The function $f^A$ has the following properties:\\
\begin{minipage}{0.48\linewidth}
\begin{enumerate}[label=(\arabic*)]
\item $f^A(\varphi+\pi) = f^A(\varphi)$
\item $f^A(-\varphi) = f^A(\varphi)$
\item $f^A(\frac{\pi}{2}+\varphi) = f(\frac{\pi}{2}-\varphi)$
\end{enumerate}
\end{minipage}
\begin{minipage}{0.51\linewidth}
\begin{enumerate}[label=]
\item $\pi$-periodic
\item symmetric, major axis
\item symmetric around $\frac{\pi}{2}$, minor axis
\end{enumerate}
\end{minipage}
\begin{minipage}{\linewidth}
	\centering
	\centering\includegraphics[width=0.7\linewidth]{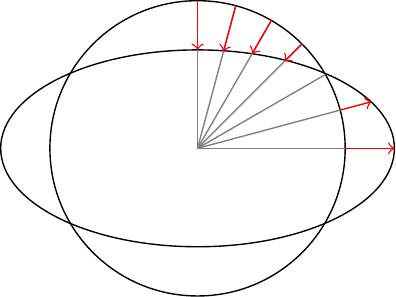}
\end{minipage}\\

Using these properties, we can derive the symmetries of the overlap reduction functions, if we define them as follows:
\begin{align}\label{eq.: FalseGamma}
&\gamma^M(\sigma) = \frac{1}{4\pi}\int_{\mathbb{S}^2} F^{M_1}_a(\hat{\Omega})F^{M_1}_b(\hat{\Omega}) + F^{M_2}_a(\hat{\Omega})F^{M_2}_b(\hat{\Omega}) d\hat{\Omega} \notag \\
&= \frac{1}{4\pi}\int_{\mathbb{S}^2} F^{M_1}_a(\hat{\Omega})F^{M_1}_b(\hat{\Omega}) d\hat{\Omega} + \frac{1}{4\pi}\int_{\mathbb{S}^2} F^{M_2}_a(\hat{\Omega})F^{M_2}_b(\hat{\Omega}) d\hat{\Omega} \notag \\
&\eqcolon \gamma^{M_1}(\sigma) + \gamma^{M_2}(\sigma),\\
&\gamma^{M_i}(\sigma) = \frac{1}{4\pi}\int_0^\pi \sqrt{1-(\hat{n}\cdot\hat{p}_a)^2}\sqrt{1-(\hat{n}\cdot\hat{p}_b)^2}\sin\theta d\theta \notag \\
&\qquad\qquad\cdot \int_0^{2\pi} f^{M_i}(\varphi)f^{M_i}(\varphi+\sigma) d\varphi,
\end{align}
where $M=\begin{pmatrix} M_1 \\ M_2 \end{pmatrix} \in \left\lbrace T= \begin{pmatrix} + \\ \times \end{pmatrix}, V= \begin{pmatrix} x \\ y \end{pmatrix}, S= \begin{pmatrix} b \\ l \end{pmatrix} \right\rbrace$ are the tensor $T$, vector $V$ and scalar $S$ polarization modes, respectively.\\

We can show, that the symmetries of the polarizations carry over to the overlap reduction function by proving, that the $\varphi$-integral $I(\sigma)$ is symmetric around $\sigma = \frac{\pi}{2}$, using the properties of $f^A$, since the projection onto the ellipse plane does not depend on $\sigma$:
\begin{align}
I\left(\frac{\pi}{2}+\sigma\right) &\coloneq \int_0^{2\pi} f^A(\varphi)f^A\left(\varphi+\frac{\pi}{2}+\sigma\right) d\varphi \notag \\
&\overset{(3)}{=} \int_0^{2\pi} f^A(\varphi)f^A\left(\frac{\pi}{2}-\varphi-\sigma\right) d\varphi \notag \\
&\overset{\varphi\mapsto-\varphi}{=} -\int_0^{-2\pi} f^A(-\varphi)f^A\left(\varphi+\frac{\pi}{2}-\sigma\right) d\varphi \notag \\
&= \int_{-2\pi}^0 f^A(-\varphi)f^A\left(\varphi+\frac{\pi}{2}-\sigma\right) d\varphi \notag \\
&\overset{(1)}{=} \int_0^{2\pi} f^A(-\varphi)f^A\left(\varphi+\frac{\pi}{2}-\sigma\right) d\varphi \notag \\
&\overset{(2)}{=} \int_0^{2\pi} f^A(\varphi)f^A\left(\varphi+\frac{\pi}{2}-\sigma\right) d\varphi \notag \\
&= I\left(\frac{\pi}{2}-\sigma\right).
\end{align}

Therefore we conclude, that the overlap reduction functions for all modes, especially the Hellings \& Downs curve (T-mode), should be symmetric around $\frac{\pi}{2}$:
\begin{equation}
\gamma^M\left(\frac{\pi}{2}+\sigma\right) = \gamma^M\left(\frac{\pi}{2}-\sigma\right),	\textit{ $ $ $ $ }	\forall M \in \lbrace T,V,S\rbrace.
\end{equation}

\begin{figure}[h!]
	\centering\includegraphics[width=\linewidth]{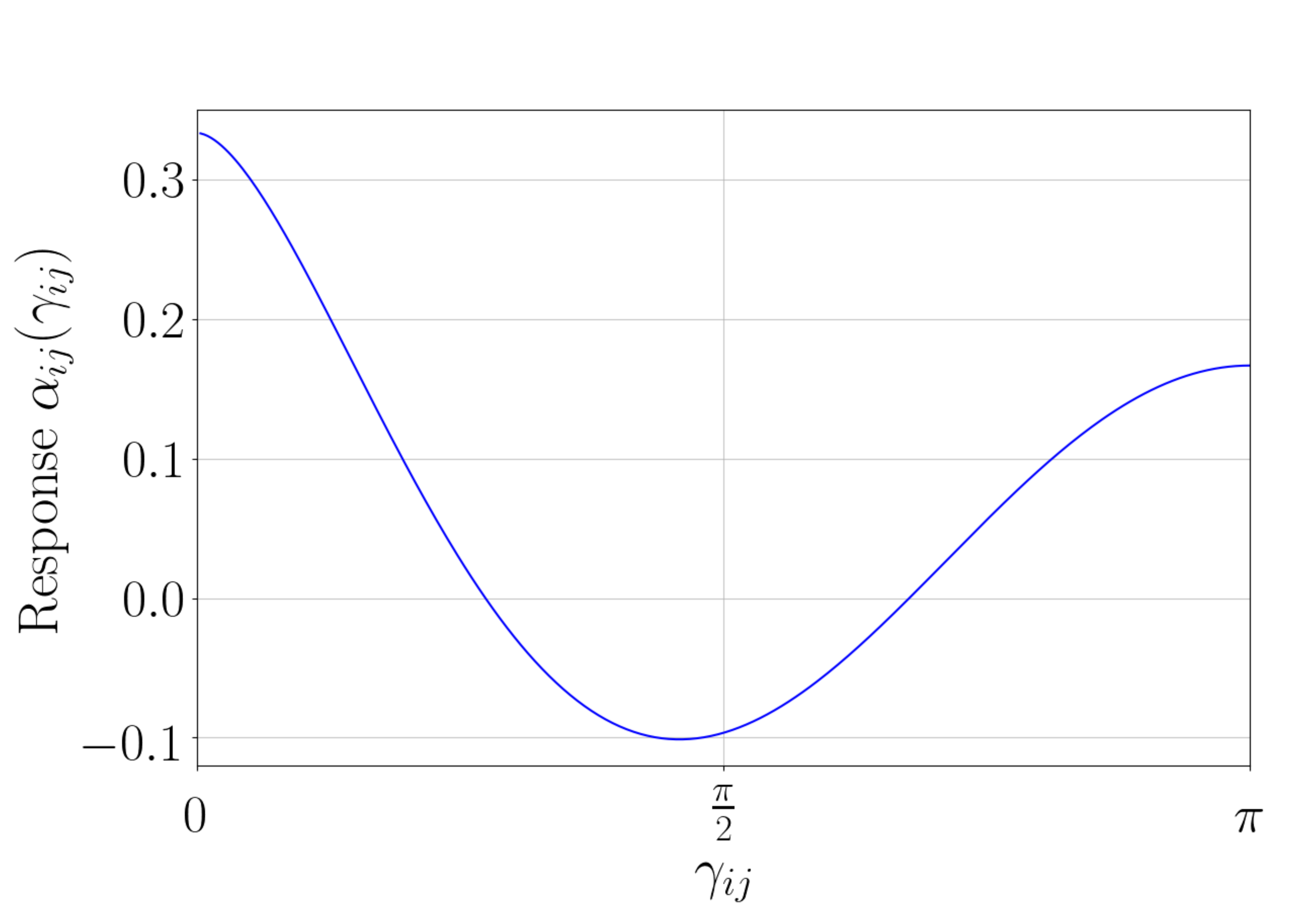}
	\caption{Hellings \& Downs curve $\alpha_{ij}(\gamma_{ij})$ which is the overlap 		
		reduction function for the tensor mode calculated with common pattern 
		functions. The angle between the pulsars $i$ and $j$ is denoted by 
		$\gamma_{ij}$}
	\label{fig: H&D}
\end{figure}

The Hellings \& Downs curve does not have this symmetry (Fig.~\ref{fig: H&D}).
\\

The definition in~\eqref{eq.: FalseGamma} is made under the assumption, that the exponential terms $e^{\mathrm{i}\frac{L\omega}{c}(1+\hat{p}^i\hat{\Omega}_i)}$ are insignificant and using the canonical pattern functions. Thus, it looks as if it would just be an integral over pattern functions and thus only the polarization part is involved. In that case the above argument would hold, and the overlap reduction function would have to satisfy this symmetry. However, the pattern functions are not the full geometric part of the signal, also the interference term has geometrical parts in it. This interference part is exactly what breaks the symmetry. We will derive the overlap reduction function for the tensorial mode in a comprehensive way in our next paper.\\

\section{The Detector Tensor}\label{sec: DetectorTensor}
A gravitational wave $h$ produces a scalar signal in a detector $D$, so the interaction of the detector with a gravitational wave is a map from the space of waves to scalar functions dependent on time:
\begin{equation}
I_D: \lbrace h_{\mu\nu} \in \mathcal{T}^2_2(\mathbb{R}^4) | \square h_{\mu\nu} = T_{\mu\nu} \rbrace \rightarrow C^\infty(\mathbb{R}).
\end{equation}

A gravitational wave in modified general relativity in the most general case can be written as \cite{Will2014}:
\begin{equation}
h_{ij}(t,\hat{\Omega}) = \sum_A h_A(t,\hat{\Omega})e_{ij}^A(\hat{\Omega}),
\end{equation}
with the amplitude $h_A$ of the polarization $A$ and $e^A$ the unit polarization tensor.\\

In the interaction with the detector, two things can happen: The waveform can induce an interference in the detector and the detector arms can point in a direction in which a specific polarization distorts space or not.\\
The first can be described by an arbitrary scalar function $f$ of the wave form $h_A$ and the letter by contracting the polarization tensors with the detector tensor $D$, which is the sum of tensor products of unit direction vectors $\hat{v}_n$ of all detector arms:
\begin{align}
I_D[h] = \sum_A f[h_A(t,\hat{\Omega})] D^{ij}e^A_{ij}(\hat{\Omega}), && D^{ij} \coloneq \frac{1}{2}\sum_{n=1}^N \hat{v}_n \otimes \hat{v}_n.
\end{align}
Since the wave forms $h_A$ can in general be arbitrary functions of time $t$ and direction $\hat{\Omega}$ for any polarization, the function $f$ cannot depend on the polarization.\\
A specific polarization influences at most two independent space directions. Thus, the full signal can be captured, by having a detector arm point in each of the two directions. Therefore, the detector tensor is normalized with the prefactor $\frac{1}{2}$, to reflect the fact, that we can at most measure the full signal or nothing and thus the absolute value of the pattern functions must lie between $0$ and $1$.\\

So, the problem is split into a part dependent on the amplitude $h_A$ and one on the polarization. The pattern functions capture the polarization part and their definition is therefore the contraction of the polarization tensor with the detector tensor:
\begin{equation}
F^A_D(\hat{\Omega}) \coloneq D^{ij}e_{ij}^A(\hat{\Omega}).
\end{equation}\\

A PTA is essentially a detector with one arm. Therefore, the detector tensor is given by:
\begin{equation}
D = \frac{1}{2}\hat{p}\otimes\hat{p},
\end{equation}
with $\hat{p}$ being the direction in which the pulsar is located.\\
\\
We can describe the orthonormal basis $\hat{\Omega}, \hat{m}, \hat{n}$ of the GW reference frame, as introduced in~\cite{Nishizawa2009}, and the pulsar location $\hat{p}$ in a Cartesian reference frame and write the pattern functions as scalar products of these unit vectors, using the following expressions for the polarization tensors, which satisfy $e^A_{ij}e_{A'}^{ij} = 2\delta_{AA'}$:
\begin{align}
	e^+ &= \hat{m}\otimes\hat{m} - \hat{n}\otimes\hat{n},	&&	e^{\times} = \hat{m}\otimes\hat{n} + \hat{n}\otimes\hat{m}, \notag \\
	e^x &= \hat{m}\otimes\hat{\Omega} + \hat{\Omega}\otimes\hat{m},	&&	e^y = \hat{n}\otimes\hat{\Omega} + \hat{\Omega}\otimes\hat{n}, \notag \\
	e^b &= \hat{m}\otimes\hat{m} + \hat{n}\otimes\hat{n}, &&	e^l = \sqrt{2}\hat{\Omega}\otimes\hat{\Omega}.
\end{align}

Defining the direction cosines of the pulsar in the GW frame as:
\begin{align}
\alpha &\coloneq \cos\theta_m = \hat{m}\cdot\hat{p}, && \beta \coloneq \cos\theta_n = \hat{n}\cdot\hat{p}, \notag \\
\gamma &\coloneq \cos\theta_\Omega = \hat{\Omega}\cdot\hat{p},
\end{align}
we can write the pattern functions as:
\begin{align}\label{eq: FA}
F^+ &= \frac{\alpha^2 - \beta^2}{2}, && F^x = \alpha\gamma, && F^b = \frac{\alpha^2 + \beta^2}{2}, \notag \\
F^\times &= \alpha\beta,  && F^y = \beta\gamma, && F^l = \frac{\gamma^2}{\sqrt{2}}.
\end{align}

\begin{figure}[h!]
	\centering\includegraphics[width=\linewidth]{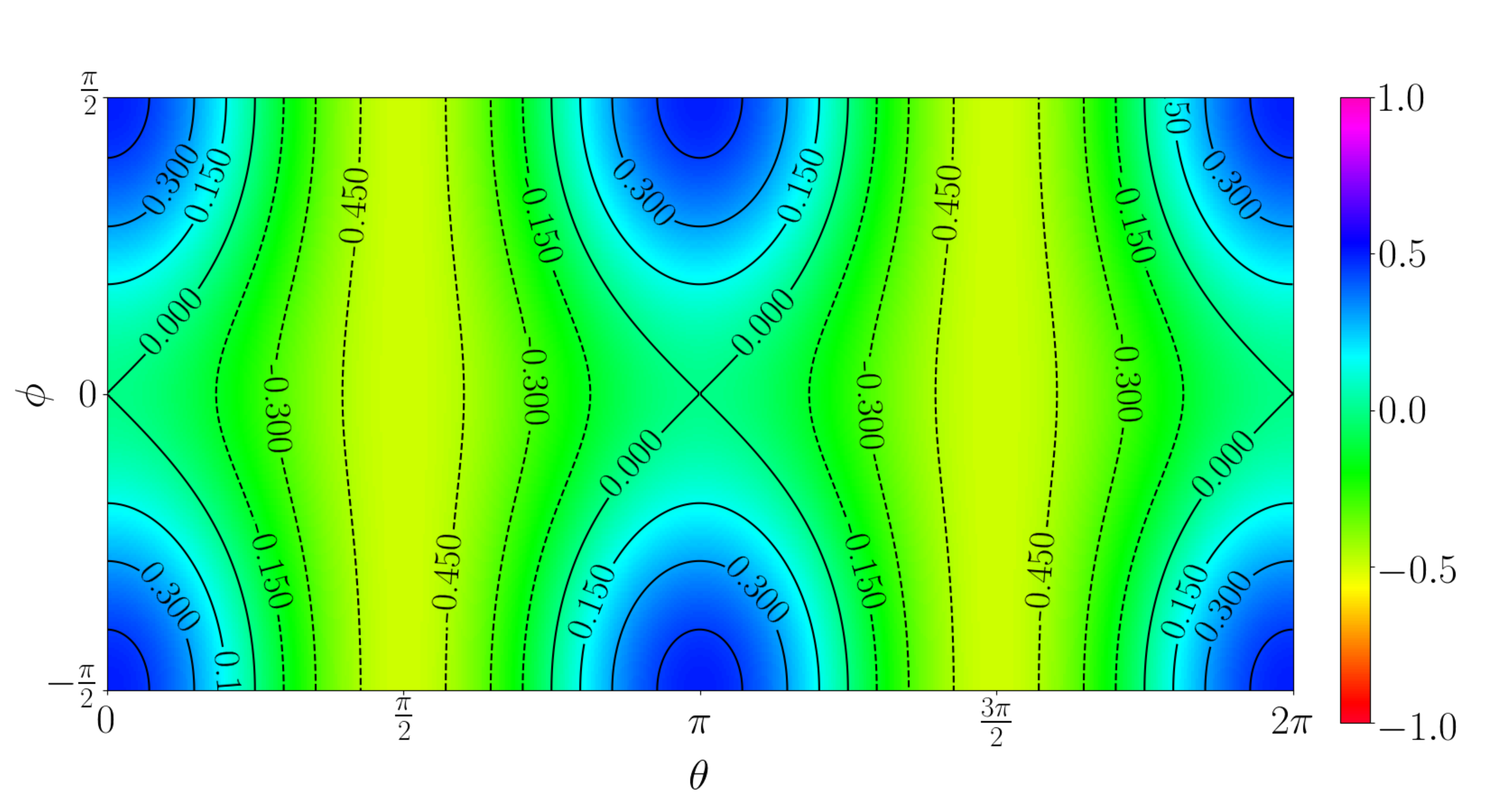}
\caption{The angular pattern function for the $+$ polarization $F^+(\hat{\Omega})$ according to the above definition.}
\label{fig: F+(Omega)-smooth}
\end{figure}

We plot the redefined pattern function (in accordance to interferometers) for the $+$ polarization in Fig.~\ref{fig: F+(Omega)-smooth} for comparison with the original one in Fig.~\ref{fig: F+(Omega)}.

\section{Distinction between photon and pulse redshift}\label{sec: diffRedshifts}
The redshift of a photon can be defined and rewritten as follows:
\begin{align}\label{eq: z}
	z &\coloneq \frac{E_\mathcal{E}-E_\mathcal{R}}{E_\mathcal{E}} = \frac{E_\mathcal{E}}{E_\mathcal{R}}-1
	= \frac{h\nu_\mathcal{E}}{h\nu_\mathcal{R}}-1 \notag \\
	&\overset{\nu=\frac{1}{T}}{=} \frac{T_\mathcal{R}}{T_\mathcal{E}}-1
	= \frac{T_\mathcal{R}-T_\mathcal{E}}{T_\mathcal{E}} = \frac{\Delta T}{T_\mathcal{E}},
\end{align}
where $\mathcal{E}$ denotes the emitter (pulsar) and $\mathcal{R}$ the receiver (Earth), the energy of a photon is given by $E = h\nu$ with Planck's constant $h$ and the period $T$ is the inverse of the frequency $\nu = \frac{1}{T}$.\\

In the case of a single photon we use the energy $E$, which is the eigenvalue of the Hamiltonian operator as an observable to assign a frequency to the quantum particle, which corresponds to the frequency of its wave function.\\
While in the case of many photons where the limit of a continuous electromagnetic wave is achieved, the wave function becomes a real wave and thus we use period T to compute the frequency.\\ 

In the case of PTA's however, we are not interested in energy, frequency or period of the photon. We are also not interested in the period of the pulses at the pulsar ($\mathcal{E}$) compared to the ones at Earth ($\mathcal{R}$), as equation~\ref{eq: z} would suggest. We want to know, how the measured period of a pulsar changes, when a gravitational wave is present, compared to the case without.
\begin{align}
	z_P &\coloneq \frac{\Delta T}{T_a} \overset{i.g.}{\neq} \frac{\Delta T}{T_\mathcal{E}}: \quad \Delta T = t_{\mathcal{R},\gamma_2} - t_{\mathcal{R},\gamma_1} - T_a, \notag \\
	T_\mathcal{R} &= t_{\mathcal{R},\gamma_2} - t_{\mathcal{R},\gamma_1}, \quad \text{period measured at Earth} \notag \\
	 T_a &= t_{\mathcal{E},\gamma_2} - t_{\mathcal{E},\gamma_1}|_{h=0} \overset{\text{i.g.}}{\neq} T_\mathcal{E} = t_{\mathcal{E},\gamma_2} - t_{\mathcal{E},\gamma_1}|_{h\overset{i.g.}{\neq}0},
\end{align}
with pulsar period $T_a$ and the two photons $\gamma_{1,2}$ of subsequent pulses.

\section{Pulse Redshift}\label{sec: PulseRedshift}
When a gravitational wave passes through our galaxy it changes the distance between Earth and a pulsar $a$. This causes a redshift in the photon frequency and also in the frequency of the pulses arriving at Earth. Pulsar timing array (PTA) experiments measure the redshift in the pulses of a collection of photons to detect a gravitational wave.\\

Our method to calculate this redshift to any desired order works as follows:\\
First, we need to calculate the geodesic of a photon that leaves the pulsar and arrives at earth (sec.~\ref{sec. P.Geod.}). One starts at the 0-th order (sec.~\ref{sec. 0-ord}) and then solves the initial value problems order by order, using the solutions of the previous orders. We stop at the first order (sec.~\ref{sec. 1-ord}).\\
Now we can proceed to calculate the arrival time difference (sec.~\ref{sec. a.t.diff.}), where we need to take into account, that since the angles change due to the presence of a GW two subsequent photons, which hit earth, leave the pulsar in slightly different directions and thus the emission time difference deviates from the pulsar period. To compare the initial momenta at the two subsequent emission events, we need to parallel transport them into the same tangent space (sec.~\ref{sec. PT}). Only the part which is parallel to the pulsars rotation plane contributes to the deviation from the period and thus we project the momenta onto that plane in sec.~\ref{sec. projection}. Assuming, that the pulsar rotates with constant angular velocity, the deviation is proportional to the angle between the two projected momenta. To calculate this angle we introduce a cross product which takes the curvature of space-time into account (sec.~\ref{sec. ang.dev.}). With all ingredients ready we can plug our results into the formula for the pulse redshift. We then generalize the expression to arbitrary wave forms and then expand in $\omega T_a$ (sec.~\ref{sec. redshift O(h)}), which is small for millisecond pulsars and for the frequency band in which the PTA experiments are currently measuring.\\
The major steps of our procedure are as follows:

\renewcommand{\labelenumi}{\Alph{enumi}.}
\renewcommand{\labelenumii}{\arabic{enumii}.}
\begin{enumerate}
    \item Photon Geodesics
    \begin{enumerate}
        \item 0-th order problem
        \item 1-st order perturbation
    \end{enumerate}
    \item Arrival Time Difference
    \begin{enumerate}
        \item Parallel Transport along the pulsar geodesic
        \item Projection onto the rotation Plane of the Pulsar
        \item Deviation from the full Pulsar rotation $\delta\theta$
    \end{enumerate}
    \item Redshift to first order in $h$
\end{enumerate}

\subsection{Photon Geodesics}\label{sec. P.Geod.}
To derive the redshift of these pulses we need to track the photons of the pulsar jets of two subsequent pulses (when the jet points at Earth) from the pulsar to Earth and compare their arrival times. To do this, we need to solve the geodesic equations in a perturbed Minkowski metric with the boundary condition for a massless particle and initial conditions, that the photons are emitted from the pulsar $a$ at a certain parameter value $\lambda_a$ and arrive at Earth at $\lambda_E$.\\
The metric for Minkowski spacetime, perturbed by a gravitational wave moving in $z$-direction in the most general case is given by:
\begin{align}
&g = \eta + h = \begin{pmatrix}
	-c^2 & 0 & 0 & 0 \\
	0 & 1 + h_b + h_+ & h_\times & h_x \\
	0 & h_\times & 1 + h_b - h_+ & h_y \\
	0 & h_x & h_y & 1 + \sqrt{2}h_l
\end{pmatrix}, \notag \\
&h_A = h_A\cos\left(\omega\left[t-\frac{z}{c}\right]+\varphi_A\right),
\end{align}
where $\omega$ is the frequency of the gravitational wave and $\varphi_A$ the phase of the polarization $A\in\{+,\times,x,y,b,l\}$. Geodesics are curves $\bold{x}(\lambda): I\subset\mathbb{R}\to\mathcal{M}$ on the space-time with zero acceleration:
\begin{equation}
\frac{D\dot{\bold{x}}^\mu}{d\lambda} = \ddot{\bold{x}}^\mu + \Gamma^\mu_{\rho\sigma}\dot{\bold{x}}^\rho\dot{\bold{x}}^\sigma = 0.
\end{equation}
Since the photon-momentum is tangent to its geodesic, one can always choose an affine parameter $\lambda$ such that it coincides with the canonical momentum.
\begin{align}
&\text{Lagrangian:} \quad L = \frac{1}{2}g_{\mu\nu}\dot{\bold{x}}^\mu\dot{\bold{x}}^\nu, \notag \\
&\text{canonical momentum: } \bold{p}_\mu \coloneq \frac{\partial L}{\partial\dot{\bold{x}}^\mu} = g_{\mu\nu}\dot{\bold{x}}^\nu = \dot{\bold{x}}_\mu \notag \\
&\Rightarrow \quad \bold{p}^\mu = g^{\mu\nu}\bold{p}_\nu = \underbrace{g^{\mu\nu}g_{\nu\rho}}_{\delta^\mu_\rho}\dot{\bold{x}}^\rho = \dot{\bold{x}}^\mu. \vphantom{\frac{1}{2}}
\end{align}
So, the 4-momentum of the photon coincides with its 4-velocity and thus we can rewrite the geodesic equations in terms of the momentum:
\begin{equation}
\frac{D\bold{p}^\mu}{d\lambda} = \dot{\bold{p}}^\mu + \Gamma^\mu_{\rho\sigma}\bold{p}^\rho \bold{p}^\sigma = 0.
\end{equation}

We expand to linear order in $h_A\ll 1$ $\forall A$ and choose a perturbation ansatz for the geodesic:
\begin{align}
\bold{x}^\mu(\lambda)=(\bold{x}^\mu)^{(0)}(\lambda) + \delta \bold{x}^\mu(\lambda), \quad \delta \bold{x}^\mu(\lambda)\sim\mathcal{O}(h_A)\sim\mathcal{O}(h).
\end{align}

\subsubsection{0-th order problem}\label{sec. 0-ord}
At zeroth order $\mathcal{O}(h^0)$ the metric reduces to flat space-time and the photon geodesic is a straight line:
\begin{align}
g = \eta, \quad \eta_{,\mu} = 0 \quad &\Rightarrow \quad \Gamma^\mu_{\rho\sigma} = 0 \notag \\
\Rightarrow \quad \dot{\bold{p}}^\mu = 0 \quad &\Rightarrow \quad (\bold{x}^\mu)^{(0)}(\lambda) = x_0^\mu + p_0^\mu\lambda.
\end{align}

A photon travels along a null-geodesic and thus:
\begin{equation}
\bold{p}^2 = 0 \quad \Rightarrow \quad c p_0^t = \sqrt{(p_0^x)^2
	+ (p_0^y)^2 + (p_0^z)^2},
\end{equation}
since we want the photon to propagate in positive time direction.\\

The photon shall hit Earth's world line $\bold{E}$ at $\lambda_E = 0$, which is located at the spatial origin:
\begin{equation}
(\bold{x}^\mu)^{(0)}(\lambda_E) = x_0^\mu = (ct,0,0,0) = \bold{E}^\mu(t).
\end{equation}

The photon is emitted from the pulsar located at $\vec{x}_a$ at the parameter value $\lambda_a$:
\begin{align}
&\vec{\bold{x}}^{(0)}(\lambda_a) = \vec{p}_0\lambda_a = L\vec{\alpha}, \quad \vec{\alpha} = (\alpha,\beta,\gamma) \notag \\
&\Rightarrow \quad \lambda_a = \alpha^i\frac{L}{p_0^i} \quad \forall i\in\{1,2,3\} \notag \\
&\Rightarrow \quad \vec{p}_0 = -\mathcal{P}\vec{\alpha}, \quad \mathcal{P} > 0, \quad \lambda_a = -\frac{L}{\mathcal{P}} \notag \\
&\text{and} \quad p_0^t = \frac{\sqrt{\mathcal{P}^2\vec{\alpha}^2}}{c} = \frac{\mathcal{P}}{c},
\end{align}
where $L$ is the distance to the pulsar, $\vec{\alpha}$ is the unit direction vector of the pulsar $\hat{p}$, expressed in the GW reference frame and $\mathcal{P}$ is the absolute value of the 3-momentum which depends on the choice of the affine parametrization $\lambda$ and is thus a free parameter, which we do not have to fix.\\

Thus we can write the 4-momentum as:
\begin{align}
\bold{p}^\mu(\lambda) = p_0^\mu + \delta\bold{p}^\mu(\lambda), \quad p_0^\mu = \mathcal{P}\alpha^\mu, \quad \alpha^\mu = \left(\frac{1}{c},-\vec{\alpha}\right).
\end{align}

\subsubsection{1-st order perturbation}\label{sec. 1-ord}
The argument of the gravitational wave is to first order given by:
\begin{align}
	\bold{t}(\lambda) - \frac{\bold{z}(\lambda)}{c} &\approx t + \frac{\mathcal{P}}{c}\lambda + \delta \bold{t}(\lambda)
	- \frac{-\mathcal{P}\gamma\lambda + \delta \bold{z}(\lambda)}{c} \notag \\
	&= t + \frac{\mathcal{P}}{c}[1+\gamma]\lambda + \delta \bold{t}(\lambda) - \frac{\delta \bold{z}(\lambda)}{c}
\end{align}
\begin{align}
	\Rightarrow \quad \omega\left[\bold{t}-\frac{\bold{z}}{c}\right]+\varphi_A &= \varphi + \frac{\mathcal{P\omega}}{c}[1+\gamma]\lambda + \varphi_A + \mathcal{O}(h) \notag \\
	&\eqcolon \Phi_A(\lambda) + \mathcal{O}(h),
\end{align}
where $\varphi = \omega t$ serves as a reference phase and the phase-shift of the polarization $A$ relative to it is denoted by $\varphi_A$.\\

Since the gravitational wave itself is of order $h$ we only need the argument at zeroth order:
\begin{align}
h_A\cos\left(\Phi_A(\lambda) + \delta \bold{t}(\lambda) - \frac{\delta \bold{z}(\lambda)}{c}\right)
	\approx h_A\cos\Phi_A(\lambda) + \mathcal{O}(h^2).
\end{align}

By inserting the perturbation ansatz and the zeroth order solution into the geodesic equations we obtain differential equations for the momentum perturbations:
\begin{equation}
\delta\dot{\bold{p}}^\mu \approx -\mathcal{P}\Gamma^\mu_{\rho\sigma}\alpha^\rho\alpha^\sigma
	= \frac{\mathcal{P}^2\omega}{c} I^\mu F^{\mu A} h_A \sin\Phi_A(\lambda),
\end{equation}
where we defined generalized pattern functions $F^{\mu A}$ and the prefactor $I^\mu$ which is the same for all polarizations and thus independent of it:
\begin{widetext}
	\begin{align}
		& I^t = \frac{1}{c}, &\quad &F^{tA} = F^A \quad \forall A\in\{+,\times,x,y,b,l\}, \\
		& I^x = I^y = -[1+\gamma], &\quad F^{xb} = F^{x+} &= \alpha, \quad F^{x\times} = \beta, \quad F^{xx} = \gamma, \quad F^{xy} = F^{xl} = 0, \notag \\
			& &\quad F^{yb} = -F^{y+} &= \beta, \quad F^{y\times} = \alpha, \quad F^{yy} = \gamma, \quad F^{yx} = F^{yl} = 0, \notag \\
		& I^z = 1, &\quad F^{zA} = F^A \quad &\forall A\in\{b,+,\times\}, \quad F^{zx} = -\alpha, \quad F^{zy}
			= -\beta, \quad F^{zl} = -\frac{\gamma(2+\gamma)}{\sqrt{2}}, \notag
	\end{align}
	and we use the sum convention for $A$ to denote a sum over all six polarizations.\\

	This equation can be directly integrated to obtain the 4-momentum:
	\begin{align}
		\bold{p}^\mu(\lambda) &= p_0^\mu + \delta p_0^\mu + \int_0^\lambda \delta\dot{\bold{p}}^\mu(\eta) d\eta
			= p_0^\mu + \delta p_0^\mu + \frac{\mathcal{P}^2\omega}{c}I^\mu F^{\mu A}h_A \int_0^\lambda \sin\Phi_A(\eta) d\eta \notag \\
		&= p_0^\mu + \delta p_0^\mu + \frac{\mathcal{P}}{1+\gamma}I^\mu F^{\mu A} h_A\left( \cos(\varphi+\varphi_A) - \cos\Phi_A(\lambda) \vphantom{\sqrt{2}}\right).
	\end{align}
	Another integration will lead us to the general form of a photon geodesic up to first order in $h$:
	\begin{align}
		\bold{x}^\mu(\lambda) &= x_0^\mu + \delta x_0^\mu + (p_0^\mu + \delta p_0^\mu)\lambda + \int_0^\lambda\int_0^\eta \delta\dot{\bold{p}}^\mu(\xi) d\xi d\eta \\
		 &= x_0^\mu + \delta x_0^\mu + (\mathcal{P}\alpha^\mu + \delta p_0^\mu)\lambda
			 + I^\mu F^{\mu A}\left\lbrace \frac{\mathcal{P}}{1+\gamma}h_A\cos(\varphi+\varphi_A)\lambda + \frac{c}{\omega[1+\gamma]^2}\Delta h_A^s(\lambda) \right\rbrace
			 \notag \\
		 &= \underbrace{x_0^\mu + \delta x_0^\mu}_\text{constant term}
		 + \underbrace{\left( \mathcal{P}\alpha^\mu + \delta p_0^\mu 
		 + \frac{\mathcal{P}}{1+\gamma}I^\mu F^{\mu A}h_A\cos(\varphi+\varphi_A) \right)\lambda}_\text{linear term}
		 + \underbrace{\frac{c}{\omega[1+\gamma]^2}I^\mu F^{\mu A}\Delta h_A^s(\lambda)}_\text{oscillatory term}, \notag
	\end{align}
\end{widetext}
where we define
\begin{align}\label{eq.: DhAs}
	&\Delta h_A^s(\lambda) = h_A\left( \sin(\varphi+\varphi_A) - \sin\Phi_A(\lambda) \vphantom{\sqrt{2}}\right) \\
	&= -2h_A\cos\left(\varphi+\frac{\mathcal{P\omega}}{2c}[1+\gamma]\lambda+\varphi_A\right)\sin\left(\frac{\mathcal{P}\omega}{2c}[1+\gamma]\lambda\right), \notag
\end{align}
which can also be written as a modulation, to simplify the expression.\\

Now we need to use boundary conditions to determine the integration constants $\delta \bold{x}^\mu(0)=\delta x_0^\mu$ and $\delta \bold{p}^\mu(0)=\delta p_0^\mu$. For the spatial components we again have the condition, that the photon is emitted from the pulsar at $\lambda_a=-\frac{L}{\mathcal{P}}$ and hits Earth at $\lambda_E=0$:
\begin{align}
\bold{x}^i(0) = x_0^i + \delta x_0^i = 0 \ \Rightarrow \ \delta x_0^i = 0, \, \text{since} \ x_0^i = 0.
\end{align}
and $\bold{x}^i(\lambda_a) = L\alpha^i$ determines the $\delta p_0^\mu$:
\begin{align}
	\delta p_0^i =& -\frac{\mathcal{P}}{1+\gamma}I^iF^{iA}h_A\cos(\varphi+\varphi_A) \notag \\
	&+ \frac{\mathcal{P}c}{L\omega[1+\gamma]^2}I^iF^{iA}\Delta h_A^s(\lambda_a).
\end{align}
The spatial parts of the momentum and geodesic are thus given by:
\begin{widetext}
	\begin{align}
		&\bold{p}^i(\lambda) = -\mathcal{P}\left( \alpha^i - \frac{I^i}{1+\gamma}F^{iA}
			\left\lbrace\vphantom{\frac{1}{2}}\right.\right. \underbrace{\frac{c}{L\omega[1+\gamma]}\Delta h_A^s(\lambda_a)}_\text{constant shift}
			- \underbrace{h_A\cos\Phi_A(\lambda)}_\text{oscillation} \left.\left.\vphantom{\frac{1}{2}}\right\rbrace \right), \notag \\
		&\bold{x}^i(\lambda) = -\mathcal{P}\left(\vphantom{\frac{1}{2}}\right. \alpha^i
			- \underbrace{\frac{cI^i}{L\omega[1+\gamma]^2}F^{iA}\Delta h_A^s(\lambda_a)}_\text{linear term} \left.\vphantom{\frac{1}{2}}\right)\lambda
			+ \underbrace{\frac{cI^i}{\omega[1+\gamma]^2}F^{iA}\Delta h_A^s(\lambda)}_\text{oscillatory term}.
	\end{align}
	Next, we use the null-line condition $\bold{p}^2(\lambda) = (\eta_{\mu\nu}+h_{\mu\nu}(\lambda))(p_0^\mu+\delta\bold{p}^\mu(\lambda))(p_0^\nu+\delta\bold{p}^\nu(\lambda)) = 0 \notag$ to solve for $\delta p_0^t$:
	\begin{align}
		\delta p_0^t =& -\frac{\mathcal{P}}{1+\gamma}I^tF^{tA}h_A\cos(\varphi+\varphi_A)
		+ \frac{\mathcal{P}c(2+\gamma)}{L\omega[1+\gamma]^2}I^tF^{tA}\Delta h_A^s(\lambda_a).
	\end{align}
	So, now we can write down the time component of the photon momentum:
	\begin{equation}
		\bold{p}^t(\lambda) = \mathcal{P}\left( \frac{1}{c} + \frac{I^t}{1+\gamma}F^{tA}\left\lbrace\vphantom{\frac{1}{2}}\right.\right.
			\underbrace{\frac{c(2+\gamma)}{L\omega[1+\gamma]}\Delta h_A^s(\lambda_a)}_\text{constant shift} - \underbrace{h_A\cos\Phi_A(\lambda)}_\text{oscillation} 
			\left.\left.\vphantom{\frac{1}{2}}\right\rbrace \right).
	\end{equation}

	To fix $\delta t_0$ we choose the time coordinate, such that the polarization $A$ of the gravitational wave has the phase $\varphi + \varphi_A$ at the event $\bold{x}^\mu(\lambda_a)$, where the photon is emitted from the pulsar.
	\begin{align}
	\Phi_A(\lambda_a) = \omega\left[\bold{t}(\lambda_a)-\frac{\bold{z}(\lambda_a)}{c} \right] + \varphi_A = \varphi + \varphi_A,
	\end{align}
	and thus
	\begin{align}
		\bold{t}(\lambda_a) &= \cancel{\frac{\varphi}{\omega}} + \delta t_0 \cancel{- \frac{L}{c}} - \frac{c}{\omega}\frac{\cancelto{1}{2}+\gamma}{[1+\gamma]^2}I^tF^{tA}\Delta h_A^s(\lambda_a)
			+ \cancel{\frac{c}{\omega[1+\gamma]^2}I^tF^{tA}\Delta h_A^s(\lambda_a)} \notag \\
		 &= \cancel{\frac{\varphi}{\omega}} - \cancel{\frac{L}{c}[1+\gamma] + \frac{L}{c}\gamma} - \cancel{\frac{I^z}{\omega[1+\gamma]^2}F^{zA}\Delta h_A^s(\lambda_a)}
		 + \cancel{\frac{I^z}{\omega[1+\gamma]^2}F^{zA}\Delta h_A^s(\lambda_a)} = \frac{\varphi}{\omega} + \frac{\bold{z}(\lambda_a)}{c} \notag \\
		\Rightarrow \quad \delta t_0 &= \frac{c}{\omega[1+\gamma]}I^tF^{tA}\Delta h_A^s(\lambda_a).
	\end{align}

	We can now write down the time component of the photon geodesic:
	\begin{equation}\label{eq.: tFlow}
		\bold{t}(\lambda) = \frac{\varphi}{\omega} + \frac{cI^t}{\omega[1+\gamma]}F^{tA}\Delta h_A^s(\lambda_a) + \mathcal{P}\left( \frac{1}{c} + \frac{c(2+\gamma)}{L\omega[1+\gamma]^2}I^tF^{tA}\Delta h_A^s(\lambda_a) \right)\lambda + \frac{cI^t}{\omega[1+\gamma]^2}F^{tA}\Delta h_A^s(\lambda).
	\end{equation}
\end{widetext}

Finally we get a family of photon geodesics $\{\bold{x}_\varphi^\mu(\lambda)\}_{\varphi\in\mathbb{R}}$, which can be considered as a flow $\bold{x}^\mu(\varphi,\lambda)$ from the pulsars world-line $a$ to Earth. The reference phase $\varphi$ at Earth ($\lambda=0$) parametrizes the collection of photon geodesics, while $\lambda$ parametrizes the individual geodesics themselves. Thus $\varphi$ can be seen as a time-like coordinate (it is in fact affine to $t = \frac{\varphi}{\omega}$) and $\lambda$ as a null-coordinate.\\
By comparing the time-like part with the spatial one we see, that if we define $J^\mu = \left((2+\gamma)I^t,\vec{I}\right)$, we can write the flow of photon geodesics as:
\begin{widetext}
	\begin{equation}\label{eq.: pGeod}
		\bold{x}^\mu(\varphi,\lambda) = x_0^\mu(\varphi) + \delta x_0^\mu(\varphi)
			+ \alpha^\mu\mathcal{P}\lambda + \frac{c}{\omega[1+\gamma]^2}F^{\mu A}
			\left[ \frac{J^\mu}{L}\Delta h_A^s(\varphi,\lambda_a)\mathcal{P}\lambda + I^\mu\Delta h_A^s(\varphi,\lambda) \right],
		\end{equation}
		with $x_0^t = \frac{\varphi}{\omega}$, $\vec{x}_0 = 0$ and the photon momenta, its tangent vector field, as:
		\begin{equation}
		\bold{p}^\mu(\varphi,\lambda) = \mathcal{P}\left( \alpha^\mu + \frac{F^{\mu A}}{1+\gamma}\left\lbrace \frac{cJ^\mu}{L\omega[1+\gamma]}\Delta h_A^s(\varphi,\lambda_a) 
			- I^\mu h_A\cos\Phi_A(\varphi,\lambda) \right\rbrace \right).
		\end{equation}
\end{widetext}

\subsection{Arrival Time Difference}\label{sec. a.t.diff.}
The pulse redshift is the change of the pulse frequency in presence relative to the case in absence of a GW:
\begin{equation}
z_P=\frac{\text{$\Delta $T}}{T_a}=\frac{t_{\text{obs},\gamma _2}-t_{\text{obs},\gamma_1}-T_a}{T_a},
\end{equation}
where $t_{obs,\gamma_i}$ are the arrival times (at Earth) of the first $\gamma_1$ and second photon $\gamma_2$ and $T_a$ is the period of pulsar a.\\

In fact, our solution does not only describe one geodesic but with the freedom of shifting $t_0$ it describes a family of solutions for photons leaving the pulsar at any moment. Of course, we can evaluate the solution at $t_0 + T_a$ and get the geodesic of the photon that left the pulsar exactly one pulsar period after the first photon of the jet left. However, this does not necessarily correspond to a photon in the jet which hits Earth, since in general the jet has to point slightly away from Earth such that the photons of this jet reach Earth. Thus, we need to find the next time the pulsar points towards Earth in the sense of a curved spacetime.
Since the distortion of space-time is small ($h$), this can be done by keeping track of how much the jet has to point in the "wrong" direction the first time and after a rotation:
\begin{align}
&\theta _1=0+\delta \theta _1, \quad \theta _2=2 \pi +\delta \theta _2, \quad \delta \theta =\delta \theta _1-\delta \theta _2 \\
&\Rightarrow \quad
	T=t_{\text{em},\gamma_2}-t_{\text{em},\gamma _1}=\frac{T_a \left(\theta _1-\theta _2\right)}{2 \pi }=T_a
   \left(1+\frac{\delta \theta }{2 \pi }\right), \notag
\end{align}
Both angular deviations $\delta\theta_1$ and $\delta\theta_2$ have to be calculated in the same tangent space to be comparable. The two spatial momenta of the photons $\gamma_1$ and $\gamma_2$ lie in two different tangent spaces since they were emitted at different times from the same object (pulsar a), which has the same spatial coordinates at both times. So, we parametrize the geodesic of the pulsar with $\tau$:
\begin{equation}
\bold{a}^\mu(\tau) = (c\tau,L\alpha,L\beta,L\gamma), \quad \bold{u}^\mu = \frac{d\bold{a}^\mu}{d\tau} = (c,0,0,0),
\end{equation}
which satisfy the geodesic equations.
\begin{align}
	\dot{\bold{u}}^\mu &= \frac{d\bold{u}^\mu}{d\tau} = -\Gamma^\mu_{\rho\sigma}\bold{u}^\rho\bold{u}^\sigma = c^2\Gamma^\mu_{tt}
	= c^2g^{\mu\nu}(2\cancelto{0}{g_{\nu t,t}} - \cancelto{0}{g_{tt,\nu}}), \notag \\
	g_{tt} &= -c^2, \qquad g_{ti} = 0.
\end{align}
We choose to calculate $\delta\theta$ in the tangent space at $\bold{a}(\tau_2)$ (emission of $\gamma_2$) and thus have to parallel transport $\vec{p}_{\gamma_1} = \bold{p}^i(\varphi_1,\lambda_a)$ from $\tau_1$ to $\tau_2$:
\begin{equation}
P_{\bold{a}, \tau_1, \tau_2}: T_{\bold{a}(\tau_1)}\mathcal{M} \to T_{\bold{a}(\tau_2)}\mathcal{M}
\end{equation}

\begin{figure}
	\centering\includegraphics[width=\linewidth]{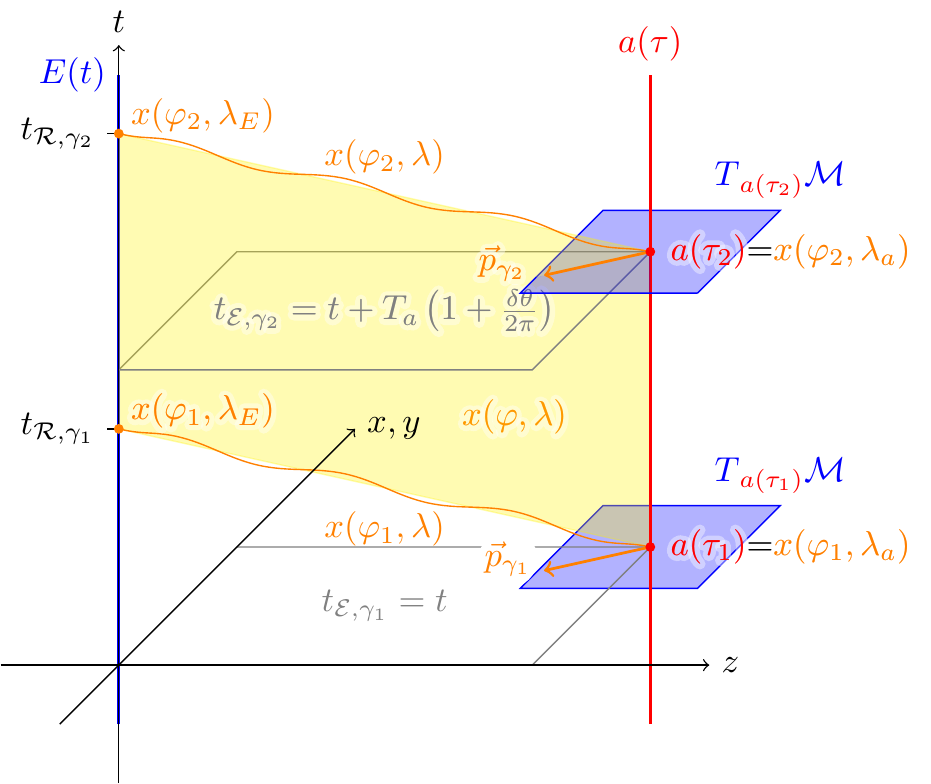}
	\caption{A sketch of the photon flow from the pulsar worldline to Earths worldline at the spatial origin with the two spatial slices at the emission of two subseqent photons. The orange vectors are the spatial comonents of the photon momenta at the emission and the spatial part of their tangent spaces at these events $\bold{a}(\tau_{1,2})$ are drawn in blue.}
	\label{fig: setup}
\end{figure}

A sketch of the setup is shown in Figure~\ref{fig: setup}.

\subsubsection{Parallel Transport along the pulsar geodesic}\label{sec. PT}
The geodesic of the pulsar is given by: $\bold{a}^\mu(\tau) = (\tau,L\vec{\alpha})$.
To find a parallel vector field $\mathcal{V}$ along $\bold{a}^\mu(\tau)$ we solve the differential equation:
\begin{equation}\label{eq.: Pfield}
\frac{D\mathcal{V}^\mu}{d\tau} = \dot{\mathcal{V}}^\mu + \Gamma^\mu_{\rho\sigma}\dot{\bold{a}}^\rho\mathcal{V}^\sigma = 0.
\end{equation}
We expand around the zeroth order case of the parallel transported 3-momentum vector:
\begin{align}
(\dot{\mathcal{V}}^\mu)^{(0)} = 0 \quad \Rightarrow \quad (\mathcal{V}^\mu)^{(0)} \eqcolon V_0^\mu = \text{const.},
\end{align}
since all Christoffel symbols are first order. Demanding, that at the event $\bold{a}(\tau_1)$ the zeroth order of the parallel field $\mathcal{V}^{(0)}$ coincides with the zeroth order momentum gives us the zeroth order vector field we are looking for:\\
\begin{align}
\mathcal{V}_0^\mu(\tau_1) = (\bold{p}^\mu)^{(0)}(\varphi_1,\lambda_a) \quad \Rightarrow \quad V_0^\mu = p_0^\mu = \mathcal{P}\alpha^\mu.
\end{align}
Inserting this solution with the linearized ansatz $\mathcal{V} = \mathcal{V}^{(0)} + \delta\mathcal{V}$ into the parallel field equation~\eqref{eq.: Pfield} results in a differential equation for the perturbation of the parallel field $\delta\mathcal{V}$.
\begin{align}
	\delta\dot{\mathcal{V}}^\mu &\approx -\Gamma^\mu_{\rho\sigma}\bold{u}^\rho\mathcal{V}^\sigma = -\frac{\mathcal{P}\omega}{2}\mathcal{F}^{\mu A}h_A\sin\Phi_A(\tau), \notag \\
	\bold{u}^\mu &= \frac{d\bold{a}^\mu}{d\tau},
\end{align}
where we use a new set of generalized pattern functions:\footnote{We note that only the z-coordinate (direction of travel of the GW) has different pattern functions and that there only the polarizations with a longitudinal part contribute.}
\begin{widetext}
	\begin{align}
		\mathcal{F}^{tA} &= 0, \quad \mathcal{F}^{xA} = F^{xA}, \quad \mathcal{F}^{yA} = F^{yA} \quad \forall A, \notag \\
		\mathcal{F}^{zA} &= 0 \quad \forall A\in\{b,+,\times\}, \quad \mathcal{F}^{zx} = \alpha, \quad \mathcal{F}^{zy} = \beta,
			\quad \mathcal{F}^{zl} = \sqrt{2}\gamma
	\end{align}
\end{widetext}
and the phase at the pulsar:
\begin{align}\label{eq: Phi(tau)}
	\Phi_A(\tau) &= \Phi_A(\lambda_a)|_{t\to\tau+\frac{L}{c}}
		= \omega\left(\tau+\frac{L}{c}\right) - \frac{L\omega}{c}[1+\gamma] + \varphi_A \notag \\
		&= \omega\left(\tau-\frac{L\gamma}{c}\right) + \varphi_A,
\end{align}
since the events $\bold{x}^\mu(\lambda_a)$ are the ones at which the photons are emitted from the pulsar and thus the same as $\bold{a}^\mu(\tau)$, so we can get the relation between $\tau$ and $t$ by inserting the two expressions into the phase of the GW:
\begin{align}
	&\bold{a}^t(\tau) - \frac{\bold{a}^z(\tau)}{c} = \tau - \frac{L\gamma}{c} = t - \frac{L}{c} - \frac{L\gamma}{c}
	= \bold{t}(\lambda_a) - \frac{\bold{z}(\lambda_a)}{c} \notag \\
	&\Rightarrow \quad \tau = t - \frac{L}{c}.
\end{align}

To get all parallel vector fields to $\bold{a}(\tau)$ we only need to integrate the sine:
\begin{align}
	&\mathcal{V}^t(\tau) = \delta V_0^t. \\
	&\mathcal{V}^i(\tau) = V_0^i + \int_{\frac{\varphi}{\omega}+\frac{L\gamma}{c}}^\tau \delta\dot{\mathcal{V}}^i(\eta) d\eta + \delta V_0^i \notag \\
	&= -\mathcal{P}\alpha^i + \delta V_0^i
	- \frac{\mathcal{P}}{2}\mathcal{F}^{iA}h_A\left(\cos(\varphi+\varphi_A)-\cos\Phi_A(\tau)\right), \notag
\end{align}
and we define $\Delta h_A^c$ in the same way as~\eqref{eq.: DhAs} to simplify the expression:\\
\begin{equation}
\Delta h_A^c(\tau) \coloneq h_A\left( \cos(\varphi+\varphi_A)  - \cos\Phi_A(\tau)\vphantom{\sqrt{2}}\right).
\end{equation}\\

If the momentum of the first photon is given by $p_{\gamma_1} = \bold{p}(\varphi_1,\lambda_a)$, then the momentum of the second photon is given by 
$p_{\gamma_2} = \bold{p}\left(\varphi_2,\lambda_a\right)$ with $\varphi_2 = \varphi_1 + \omega T_a\left(1+\frac{\delta\theta}{2\pi}\right)$.

We get the parallel transport $P_{\bold{a},\tau_1,\tau_2}(\vec{p}_{\gamma_1})$ of the first photon 3-momentum at $\tau_1$ along $\bold{a}$ to $\tau_2$, where the second photon is emitted, by choosing $\mathcal{V}^\mu(\tau_1) = (0,\bold{p}^i(\varphi_1,\lambda_a))$ as initial conditions and then evaluating $\mathcal{V}^\mu$ at $\tau_2$.\\
\begin{widetext}
	\begin{align}\label{eq.: Ptop}
		\mathcal{V}^i(\tau_1) &= \bold{p}^i(\varphi_1,\lambda_a) = -\mathcal{P}\left( \alpha^i
			- \frac{I^iF^{iA}}{1+\gamma}\left\lbrace \frac{c\Delta h_A^s(\varphi_1,\lambda_a)}{L\omega[1+\gamma]}
			- h_A\cos\Phi_A(\varphi_1,\lambda_a) \right\rbrace \right),
		\end{align}
\end{widetext}
where $\tau_1$ denotes the parameter of the pulsar geodesic at which the first photon is emitted. This can be at \textit{any value} of $\varphi_{1}$, which we thus henceforth denote with $\varphi$.\\
\begin{align}
&\bold{a}^t(\tau_1) = \tau_1 = t_a^{(0)} + \delta t_a = \bold{t}(\varphi,\lambda_a) \ \Rightarrow \ \tau_1 \overset{\mathcal{O}(h)}{\approx} \frac{\varphi}{\omega} - \frac{L}{c}, \notag \\
&\Rightarrow \quad \Phi_A(\tau_1) \approx \varphi - \frac{L\omega}{c}[1+\gamma] + \varphi_A = \Phi_A(\varphi,\lambda_a).
\end{align}

Solving this equation~\eqref{eq.: Ptop} for $\delta V_0^i$ we get:
\begin{align}
	\delta V_0^i \approx& \frac{\mathcal{P}I^iF^{iA}}{1+\gamma}\left\lbrace
	\frac{c\Delta h_A^s(\tau_1)}{L\omega[1+\gamma]} - h_A\cos\Phi_A(\tau_1) \right\rbrace \notag \\
	&+ \frac{\mathcal{P}}{2}\mathcal{F}^{iA}\Delta h_A^c(\tau_1),
\end{align}
and since we are only interested in the spatial components, we set the time component to zero:
\begin{equation}
	\mathcal{V}^t(\tau_1) = 0 \quad \Rightarrow \quad \delta V_0^t = 0.
\end{equation}

So, finally the parallel transport of $\vec{p}_{\gamma_1}$ is given by:
\begin{widetext}
	\begin{align}
		p_1^i &\coloneq P^i_{a,\tau_1,\tau_2}(\vec{p}_{\gamma_1}) = P^i_{a,\tau_1,\tau_2}(\vec{\bold{p}}(\varphi,\lambda_a)) = \mathcal{V}^i(\tau_2), 
			\qquad P^t_{a,\tau_1,\tau_2}(p^t_{\gamma_1}) = 0 \notag \\
			&= -\mathcal{P}\left( \alpha^i - \frac{I^iF^{iA}}{1+\gamma}\left\lbrace \frac{c\Delta h_A^s(\tau_1)}{L\omega[1+\gamma]}
			- h_A\cos\Phi_A(\tau_1) \right\rbrace\right.
			+ \underbrace{\frac{\mathcal{F}^{iA}}{2}\left(\Delta h_A^c(\tau_2)-\Delta h_A^c(\tau_1)
			\vphantom{\sqrt{2}}\right)}_\text{parallel transport term} \left.\vphantom{\frac{1}{2}}\right).
	\end{align}
\end{widetext}
The parallel transport term can be simplified to:
\begin{equation}
\delta p_T^i = -\frac{\mathcal{P}}{2}\mathcal{F}^{iA}\left(\cos\Phi_A(\tau_2)-\cos\Phi_A(\tau_1)\vphantom{\sqrt{2}}\right),
\end{equation}
with
\begin{align}\label{eq: tau2}
\tau_2 = \frac{\varphi}{\omega} + T_a\left(1+\frac{\delta\theta}{2\pi}\right) - \frac{L}{c}
\end{align}
and thus
\begin{align}
	\Phi_A(\tau_2) &= \omega\left(\frac{\varphi}{\omega} + T_a\left(1+\frac{\delta\theta}{2\pi}\right)
	- \frac{L}{c}-\frac{L\gamma}{c}\right) + \varphi_A \notag \\
	&\approx \varphi + \omega\left(T_a - \frac{L}{c}[1+\gamma] \right) + \varphi_A + \mathcal{O}(h).
\end{align}

\subsubsection{Projection onto the rotation Plane of the Pulsar}\label{sec. projection}
To get the fraction of the full pulsar rotation, by which the emission interval of the two photons differs, we need to project the two vectors 
$\vec{p}_1=P_{a,\tau_1,\tau_2}(\vec{p}_{\gamma_1})$ and $\vec{p}_2=\vec{p}_{\gamma_2}$ onto the rotation plane of the pulsar.\\
\begin{equation}
P_{\hat{\omega}_a}(\vec{v}) \coloneq \vec{v} - \bar{g}_{a(\tau_2)}\left(\vec{v}, \vec{\omega}_a\right)\vec{\omega}_a 
	\quad \forall\, \vec{v}\in T_{a(\tau_2)} U,
\end{equation}
where $U\subset \mathcal{M}$ is the spatial slice of the pulsar.\\

Let $\bar{g}$ denote the metric restricted to a spatial slice $U$: $\bar{g} \coloneq g|_U$.\\

Since we do this calculation at the event of the emission of the second photon, the phase is given by $\Phi_A(\tau_2)$ and thus the spatial part of the metric reads:
\begin{equation}
\bar{g}(\tau_2) = \bar{\eta} + h_A\cos\Phi_A(\tau_2)\bar{e}^A,
\end{equation}

We describe the rotation vector of the pulsar $\hat{\omega}_a$ in terms of direction cosines, just as we did with the direction of the pulsar. To construct a meaningful unit rotation vector we take the one from flat space-time $\hat{\omega}_a^{(0)} = \hat{\omega}_a|_{h=0} \eqcolon \vec{\alpha}_\omega$ and imagine that we tune up the amplitude of the GW from zero to a finite value. We are not changing the reference frame, only the metric. So, the same components are still valid however due to the changed metric the same components will now have slightly different angles with respect to other directions as before. In other words: although we have the exact same $\mathbb{R}^3$-element, it now represents a vector which points slightly in a different direction. Although it traces out the right directions, it is not a unit vector anymore and we have to normalize it again using the perturbed metric.\\
So, let
\begin{equation}
\vec{\alpha}_\omega = (\alpha_\omega,\beta_\omega,\gamma_\omega)
\end{equation}
be the unit rotation vector for $h = 0$:
\begin{align}
	\hat{\omega}_a &= \frac{\vec{\alpha}_\omega}{\Vert\vec{\alpha}_\omega\Vert} = \vec{\alpha}_\omega\left( 1 - F^A_\omega h_A \cos\Phi_A(\tau_2) \right), \notag \\
	\Vert\vec{v}\Vert &= \sqrt{\bar{g}_{a(\tau_2)}(\vec{v},\vec{v})},
\end{align}
where $F^A_\omega$ are the pattern functions of the rotation vector instead of the direction vector:
\begin{equation}
F^A_\omega \coloneq \frac{1}{2}\vec{\alpha}_\omega\otimes\vec{\alpha}_\omega\cdot e^A.
\end{equation}

We project the two vectors $\vec{p}_1$, the momentum of the first photon at its emission parallel transported to the event of the emission of the second one, and $\vec{p}_2$, the emission momentum of the second photon.\\
\begin{align}
	\vec{p}_1 &\coloneq P_{a,\tau_1,\tau_2}(\vec{p}_{\gamma_1}), \\
	\vec{p}_2 &\coloneq \vec{p}_{\gamma_2} \overset{\mathcal{O}(h)}{\approx} \vec{\bold{p}}(\varphi+\omega T_a,\lambda_a) \notag \\
	&= -\mathcal{P}\left( \alpha^i - \frac{I^iF^{iA}}{1+\gamma}\left\lbrace \frac{c\Delta h_A^s(\tau_2)}{L\omega[1+\gamma]}
	- h_A\cos\Phi_A(\tau_2) \right\rbrace \right), \notag
\end{align}
which coincide at zeroth order: $\vec{p}_i = \vec{p} + \delta\vec{p}_i$ \quad $i\in\{1,2\}$.\\

The most involved part of the projection is the scalar product of the two momentum vectors with the rotation vector:\\
\begin{align}
	&\bar{g}_{a(\tau_2)}(\vec{p}_2,\hat{\omega }_a) = -\mathcal{P}\left(\vphantom{\frac{1}{2}}\right.
		\underbrace{\vec{\alpha}\cdot\vec{\alpha }_\omega}_{\bar{\eta}(\vec{p},\vec{\alpha}_\omega)}
		+ \underbrace{2 F^A_{\{\alpha,\omega\}} h_A \cos\Phi_A(\tau_2)}_{\bar{h}(\vec{p},\vec{\alpha}_\omega)} \notag \\
	&\quad - \underbrace{\vec{\alpha}\cdot\vec{\alpha}_\omega F_\omega^A h_A\cos\Phi_A(\tau_2)}
		_{\bar{\eta}(\vec{p},\delta\hat{\omega}_a)} \notag \\
	&\quad - \underbrace{ \alpha _{\omega i}\frac{I^i F^{iA}}{1+\gamma}\left\lbrace
		\frac{c\Delta h_A^s(\tau_2)}{L\omega[1+\gamma]} - h_A\cos\Phi_A(\tau_2) \right\rbrace }
		_{\bar{\eta}(\delta\vec{p}_2,\vec{\alpha}_\omega)} \left.\vphantom{\frac{1}{2}}\right).
\end{align}
The only part that changes for $\vec{p}_1$ is:
\begin{widetext}
	\begin{align}
		\bar{\eta}(\delta\vec{p}_1,\vec{\alpha}_\omega) = \mathcal{P} \alpha_{\omega i}
		   \left( \frac{I^i F^{iA}}{1+\gamma} \left\lbrace \frac{c\Delta h_A^s(\tau_1)}{L\omega[1+\gamma]} - h_A\cos\Phi_A(\tau_1)\right\rbrace
		   + \underbrace{ \frac{\mathcal{F}^{iA}}{2} h_A \left(\cos\Phi_A(\tau_2) - \cos\Phi_A(\tau_1)\vphantom{\sqrt{2}}\right) }
		   _{\text{parallel transport term}} \right),
	\end{align}
	where we defined the mixed pattern functions
	\begin{equation}
		F^A_{\{\alpha,\omega\}} \coloneq \frac{1}{2} \vec{\alpha}\otimes\vec{\alpha}_\omega\cdot e^A
	\end{equation}
	and the up and down $i$ stand for Einstein summation over the spatial components, with $\alpha_{\omega i} = \alpha_\omega^i$.

	The last step is just inserting the expressions we derived, expanding to linear order and collecting terms:
	\begin{align}
		P_{\hat{\omega}_a}(\vec{p}_1)^i =& p_1^i - \bar{g}_{a(\tau_2)}(\vec{p}_1,\hat{\omega}_a)\hat{\omega}_a^i \notag \\
		=& - \mathcal{P}\left(\vphantom{\frac{1}{2}} \alpha^i - \alpha_\omega^i \vec{\alpha}\cdot\vec{\alpha}_\omega 
			- 2\alpha_\omega^i\left(F^A_{\{\alpha,\omega\}} - \vec{\alpha}\cdot\vec{\alpha}_\omega F_\omega^A\right) h_A\cos\Phi_A(\tau _2) \right.\notag \\
		&\qquad\quad - \frac{I^i F^{iA} - \alpha_\omega^i \alpha_{\omega j} I^j F^{jA}}{1+\gamma}\left\lbrace\frac{c\Delta h_A^s(\tau_1)}{L\omega[1+\gamma]}
			- h_A\cos\Phi_A(\tau_1)\right\rbrace \notag \\
		&\qquad\quad\left. + \frac{1}{2}\left(\mathcal{F}^{iA} + \alpha_\omega^i\alpha_{\omega j}\mathcal{F}^{jA}\right) h_A\left(\cos\Phi_A(\tau_2)
			- \cos\Phi_A(\tau_1)\vphantom{\sqrt{2}}\right)\right), \notag \\
		P_{\hat{\omega}_a}(\vec{p}_2) =& p_2^i - \bar{g}_{a(\tau_2)}(\vec{p}_2,\hat{\omega}_a)\hat{\omega}_a^i \notag \\
		=& - \mathcal{P}\left(\vphantom{\frac{1}{2}} \alpha^i-\alpha_\omega^i \vec{\alpha}\cdot\vec{\alpha}_\omega - 2\alpha_\omega^i\left(F^A_{\{\alpha,\omega\}}
			-\vec{\alpha}\cdot\vec{\alpha}_\omega F_\omega^A\right) h_A\cos\Phi_A(\tau_2) \right.\notag \\
		&\qquad\quad\left. - \frac{I^i F^{iA} - \alpha_\omega^i \alpha_{\omega j} I^j F^{jA}}{1+\gamma}
			\left\lbrace\frac{c\Delta h_A^s(\tau _2)}{L\omega[1+\gamma]} - h_A\cos\Phi_A(\tau_2)\right\rbrace\right).
	\end{align}
\end{widetext}

\subsubsection{Deviation from the full Pulsar rotation $\delta\theta$}\label{sec. ang.dev.}
Angles on a manifold are defined through the metric for example via the scalar product:\\
\begin{align}\label{eq: skp}
	g_p(X,Y) &= \Vert X\Vert \Vert Y\Vert \cos\theta, \notag \\
	\Vert X\Vert &\coloneq \sqrt{g_p(X,X)} \quad \forall X,Y \in T_p\mathcal{M}.
\end{align}
Since this is a cosine we do not get a first order equation in $h$ for $\delta\theta$:
\begin{align}
	\bar{g}_{a(\tau_2)}(\vec{p}_1,\vec{p}_2) &= g_{ij}(a(\tau_2))p_1^ip_2^j = \Vert\vec{p}_1\Vert \Vert\vec{p}_2\Vert 		\cos\delta\theta \notag \\
	&\approx \Vert\vec{p}_1\Vert \Vert\vec{p}_2\Vert \frac{\delta\theta^2}{2}, \\
	\Vert\vec{v}\Vert &\coloneq \sqrt{\bar{g}_{a(\tau_2)}(\vec{v},\vec{v})}, \quad \forall v\in T_{a(\tau_2)}U, \notag
\end{align}
where $\bar{g}=(g_{ij})_{i,j\in\{1,2,3\}}$ is the metric restricted to the spatial submanifold $U$ in our current reference frame.\\
So, we would need an equation which includes a sine instead. We can get such an equation using the cross-product.
However, the standard cross-product on the spatial 3-dimensional submanifold does not satisfy the equation we need:
\begin{equation}\label{eq: cross}
\vec{a}\times\vec{b} = \Vert \vec{a}\Vert \Vert \vec{b}\Vert \sin\theta\, \hat{n}, \quad \forall\vec{a},\vec{b}\in\mathbb{R}^3.
\end{equation}
The angle we get using this equation on $\vec{p}_1$ and $\vec{p}_2$ is not consistent with the one we get using the scalar product:
 $\sin^2\theta + \cos^2\theta = 1 + \mathcal{O}(h) \neq 1$.\\
It is not difficult to understand why this is the case. The metric encodes the curved geometry of space-time,
yet all the wedge product does, is permuting the components of the two vectors, but no component of the metric is included in
the calculation. Thus, we see, that the fact that the cross product "coincides" with the wedge product in three dimensions is
merely a coincidence which happens only in flat space and have to generalize the cross-product in a different way to be able
to consistently use it in a curved space.

We use \eqref{eq: cross} and $\sin^2\theta + \cos^2\theta = 1$ $\forall\theta\in\mathbb{R}$ as a defining property for our generalized cross-product:
\begin{align}
	\Vert\vec{a}\times\vec{b}\Vert^2 &= \Vert\vec{a}\Vert^2 \Vert\vec{b}\Vert^2 \sin^2\theta
		\cancelto{1}{\Vert\hat{n}	\Vert^2} \notag \\
	\Rightarrow \quad \sin^2\theta + \cos^2\theta &= \frac{\Vert\vec{a}\times\vec{b}\Vert^2}{\Vert\vec{a}\Vert^2\Vert\vec{b}\Vert^2} + \frac{(\vec{a}\cdot\vec{b})^2}{\Vert\vec{a}\Vert^2\Vert\vec{b}\Vert^2} = 1.
\end{align}
We define the curved space cross-product to be the bilinear map:
\begin{align}
\begin{matrix}
\mathcal{E}: & \mathcal{M} & \to & C^\infty((T_pU)^2;T_pU) $ $, \\
	& p & \mapsto & \mathcal{E}(p)
\end{matrix} \notag \\
\vphantom{a}\notag\\
\begin{matrix}
\mathcal{E}(p): & T_pU\times T_pU & \to & T_pU, \\
	& X, Y & \mapsto & \mathcal{E}_p(X,Y)
\end{matrix}
\end{align}
such that
\begin{align}
	&\bullet\quad \frac{\bar{g}_p(\mathcal{E}_p(X,Y),\mathcal{E}_p(X,Y)) + \bar{g}_p(X,Y)^2}{\bar{g}_p(X,X)\bar{g}_p(Y,Y)} = 1
		\label{eq: sinprop} \tag{i} \\
	&\bullet\quad \mathcal{E}_p(X,Y)\perp_g X, Y \label{eq: perp} \tag{ii} \\
	&\bullet\quad \mathcal{E}_p(X,Y) = X\times Y \quad \text{for} \quad g=\eta \label{eq: flatlim} \tag{iii}
\end{align}
$\forall p\in\mathcal{M}$ and $\forall X,Y\in T_pU$ with $U\subset\mathcal{M}$ a space-like submanifold of space-time.\\

The components in local coordinates are defined as:
\begin{align}
	&\mathcal{E}_p(\partial_i(p),\partial_j(p)) = \mathcal{E}^k_{ij}(p)\partial_k(p), \notag \\
	&\partial_i\in Der_p(U)\simeq T_pU,
	\quad i,j,k\in\{1,2,3\},
\end{align}
where the spacial partial derivative form a basis of the spatial tangent space $T_pU$ at the event $p$. In appendix \ref{Ap: curvedCross} we calculate them in terms of the metric components and find them to be given by:
\begin{align}
	\mathcal{E}^k_{ij} &= \frac{1}{N_{ij}} \sqrt{\bar{g}_{ii}\bar{g}_{jj}-\bar{g}_{ij}^2} \epsilon^{kmn}\bar{g}_{mi}\bar{g}_{nj}, \notag \\
	N_{ij} &= \Vert(\epsilon^{kmn}\bar{g}_{mi}\bar{g}_{nj})_k\Vert,
\end{align}
where $N_{ij}$ is a normalization factor. The curved cross-product in local coordinates at $p$ for two arbitrary tangent vectors $X, Y \in T_pU$ is thus given by:
\begin{equation}
\mathcal{E}_p(X,Y) = \mathcal{E}^k_{ij}(p)X^i(p)Y^j(p)\partial_k(p).
\end{equation}

The two projected photon momenta coincide to zeroth order and thus we write them in the following form:
\begin{align}
	\vec{P}_1 &\coloneq P_{\hat{\omega}_a}(P_{a,\tau_1,\tau_2}(\vec{p}_{\gamma_1})) \eqcolon \vec{P} + \delta\vec{P}_1,
	\notag \\
	\vec{P}_2 &\coloneq P_{\hat{\omega}_a}(\vec{p}_{\gamma_2}) \eqcolon \vec{P} + \delta\vec{P}_2,
\end{align}
where the $\delta\vec{P}_i$ are of order $\mathcal{O}(h)$.\\
We insert our metric $\bar{g}_{ij} = \delta_{ij} + h_{ij}$ and expand the curved cross-product of the two momenta to first order in $h$. It reduces to the cross-product in flat space, since they are equal to zeroth order.
\begin{align}
	\mathcal{E}(\vec{P}_1,\vec{P}_2)^k &= (\varepsilon^k_{ij}+\delta\mathcal{E}^k_{ij})(P^i + \delta P_1^i)(P^j
		+ \delta P_2^j) \notag \\
	&\approx \underbrace{\mathcal{E}^k_{ij}P^iP^j}_{=0} + \varepsilon^k_{ij}(\delta P_1^i P^j
		+ P^i\delta P_2^j) \notag \\
	&= \vec{P} \times (\delta\vec{P}_2 - \delta\vec{P}_1) \approx (\vec{P}_1\times\vec{P}_2),
\end{align}
since the cross-product remains anti-symmetric in a curved space.\\

\begin{figure}[h!]
		
		
		
	\centering\includegraphics[width=0.5\linewidth]{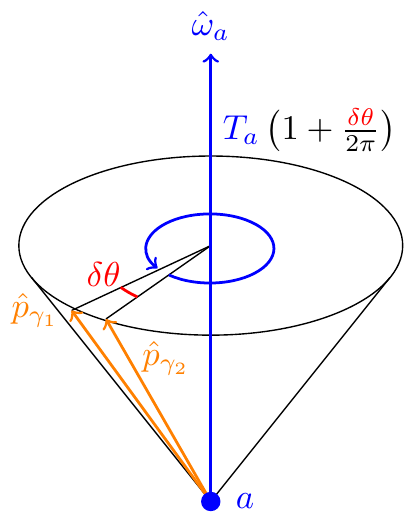}
	\caption{A projection onto the spacelike slices of the two tangent spaces $T_{\bold{a}(\tau_1)}\mathcal{M}$ and $T_{\bold{a}(\tau_2)}\mathcal{M}$. }
	\label{fig: jetcone}
\end{figure}

We apply \eqref{eq: cross} on the $\vec{P}_i$, take the scalar product with $\hat{\omega}_a$. Since both $\vec{P}_1$ and $\vec{P}_2$ lie in the rotation plane, their cross product must point in the orthogonal direction and thus $\hat{\omega}_a = \pm\hat{n}$. The zeroth order angle is zero, so $\theta = \delta\theta$. The geometry of the involved vectors is sketched in Figure~\ref{fig: jetcone}.

\begin{align}
	g(\mathcal{E}(\vec{P}_1,\vec{P}_2),\hat{\omega}_a) = \Vert\vec{P}_1\Vert \Vert\vec{P}_2\Vert \sin\theta\,
		\bar{g}\cancelto{1}{(\hat{n},\hat{\omega}_a)} \notag \\
	g(\vec{P}\times(\delta\vec{P}_2-\delta\vec{P}_1),\hat{\omega}_a)
		\approx \Vert\vec{P}_1\Vert \Vert\vec{P}_2\Vert \delta\theta.
\end{align}

Since both sides include a factor of order $h$ ($\delta\vec{P}_i$ and $\delta\theta$) all other terms contribute only at zeroth order:
\begin{align}
\left( \vec{P}\times(\delta\vec{P}_2-\delta\vec{P}_1) \right) \cdot \vec{\alpha}_\omega \approx \vert\vec{P}\vert^2 \delta\theta.
\end{align}

\begin{widetext}
	Thus we get, that the deviation angle is given by:
	\begin{align}
		\delta\theta =& \frac{1}{\vert\vec{P}\vert^2}\, \vec{\alpha}_\omega \cdot \left( \vec{P}\times(\delta\vec{P}_2-\delta\vec{P}_1)
			\right) \notag \\
		 =& -\alpha_\omega^k \varepsilon_{kij}
			 \frac{\alpha^i-\alpha_\omega^i (\vec{\alpha}\cdot\vec{\alpha}_\omega)}
				{\vert \vec{\alpha} - \vec{\alpha}_\omega (\vec{\alpha}\cdot\vec{\alpha}_\omega) \vert^2}
			\left\lbrace c\frac{I^j F^{jA} - \alpha_\omega^j \alpha_{\omega l} I^l F^{lA}}{L\omega[1+\gamma]^2}
			\left( \Delta h_A^s(\tau_2) - \Delta h_A^s(\tau_1) \vphantom{\sqrt{2}}\right) \right. \notag \\
		 &\left. + \left(\frac{1}{2}\left(\mathcal{F}^{jA} + \alpha_\omega^j \alpha_{\omega l} \mathcal{F}^{lA}\right)
			- \frac{I^j F^{jA} - \alpha_\omega^j\alpha_{\omega l} I^l F^{lA}}{1+\gamma}\right)
			h_A \left(\cos\Phi_A(\tau_2) - \cos\Phi_A(\tau_1)\vphantom{\sqrt{2}}\right) \right\rbrace,
	\end{align}
	since $\hat{n} = \hat{\omega}_a$, if we define the sign of the angle $\theta$ consistently.
\end{widetext}

\subsection{Redshift to first order in h}\label{sec. redshift O(h)}
As pointed out in \ref{sec: diffRedshifts} the pulse redshift we are looking for is given by the arrival time difference of two subsequent pulses divided by the pulsar period.\\
We know the time component of the photon flow~\eqref{eq.: tFlow} and thus we can write down the redshift at the time $t = \frac{\varphi}{\omega}$, by evaluating it at $\lambda_E=0$:\\
\begin{widetext}
	\begin{align}\label{eq.: z_P}
		z_P(t) =& \frac{\Delta T}{T_a} = \frac{t_{\mathcal{R},\gamma_2} - t_{\mathcal{R},\gamma_1} - T_a}{T_a}
			= \frac{\bold{t}(t + T_a\left(1+\frac{\delta\theta}{2\pi}\right),\lambda_E) - \bold{t}(t,\lambda_E) - T_a}{T_a} \notag \\
		\overset{\mathcal{O}(h)}{\approx}& F^A \frac{\Delta h_A^s(\tau_2) - \Delta h_A^s(\tau_1)}{\omega T_a[1+\gamma]}
			+ \frac{\delta\theta}{2\pi} \notag \\
		=& F^A \frac{\Delta h_A^s(\tau_2) - \Delta h_A^s(\tau_1)}{\omega T_a[1+\gamma]}
			- \frac{\alpha_\omega^k \varepsilon_{kij}}{2\pi}
			\frac{\alpha^i-\alpha_\omega^i (\vec{\alpha}\cdot\vec{\alpha}_\omega)}
				{\vert \vec{\alpha} - \vec{\alpha}_\omega (\vec{\alpha}\cdot\vec{\alpha}_\omega) \vert^2}
			\left\lbrace c\frac{I^j F^{jA} - \alpha_\omega^j \alpha_{\omega l} I^l F^{lA}}{L\omega[1+\gamma]^2}
			\left( \Delta h_A^s(\tau_2) - \Delta h_A^s(\tau_1) \vphantom{\sqrt{2}}\right) \right. \notag \\
		 &\left. + \left(\frac{1}{2}\left(\mathcal{F}^{jA} + \alpha_\omega^j \alpha_{\omega l} \mathcal{F}^{lA}\right)
			- \frac{I^j F^{jA} - \alpha_\omega^j\alpha_{\omega l} I^l F^{lA}}{1+\gamma}\right)
			h_A \left(\cos\Phi_A(\tau_2) - \cos\Phi_A(\tau_1)\vphantom{\sqrt{2}}\right) \right\rbrace.
	\end{align}

	If we expand in $\omega T_a$ to first order, we get:
	\begin{align}\label{eq.: z_PapproxT}
	z_P(t) \overset{\mathcal{O}(\omega T_a)}{\approx}& \frac{F^A \Delta h_A^c(\tau_1)}{1+\gamma}
		- \frac{\omega T_a}{2} \frac{F^A \Delta h_A^s(\tau_1)}{1+\gamma} \notag \\
	&- \frac{\omega T_a}{2\pi} \alpha_\omega^k \varepsilon_{kij}
		\frac{\alpha^i - \alpha_\omega^i (\vec{\alpha}\cdot\vec{\alpha}_\omega)}
			{\vert \vec{\alpha} - \vec{\alpha}_\omega (\vec{\alpha}\cdot\vec{\alpha}_\omega) \vert^2}
		\left\lbrace c\frac{I^j F^{jA} - \alpha_\omega^j \alpha_{\omega l} I^l F^{lA}}{L\omega[1+\gamma]^2}
		\Delta h_A^c(\tau_1) \right. \notag \\
	&\left. + \left(\frac{1}{2} \left(\mathcal{F}^{jA} + \alpha_\omega^j\alpha_{\omega l} \mathcal{F}^{lA}\right)
		-\frac{I^j F^{jA} - \alpha_\omega^j \alpha_{\omega l} I^l F^{lA}}{1+\gamma}\right) h_A\sin\Phi_A(\tau_1) \right\rbrace,
	\end{align}
\end{widetext}
using that
\begin{align}
	h_A\cos\Phi_A(\tau_2) =& h_A\cos\left(\varphi + \omega\left(T_a - \frac{L}{c}[1+\gamma] \right)
		+ \varphi_A\right) \notag \\
	=& h_A\cos(\Phi_A(\tau_1)+\omega T_a)\notag \\
	\approx& h_A\cos(\Phi_A(\tau_1)) + h_A\sin(\Phi_A(\tau_1))\omega T_a.
\end{align}

We can generalize our plane wave to an arbitrary waveform by using Fourier transformation.
\begin{align}
	&h(x) = \frac{1}{2}\int \tilde{h}_A(k) e^{\mathrm{i}k_\mu x^\mu + \mathrm{i}\varphi_A} + c.c.\, \frac{d^4k}{(2\pi)^4}, \notag \\
	&\text{with} \quad k^\mu = \left(\omega, \frac{\hat{\Omega}}{c}\right).
\end{align}
Without loss of generality we continue for now with a wave from a single source\\
$\tilde{h}_A(k) = \tilde{h}_A(\omega)\delta(\hat{\Omega}-\hat{\Omega}')$ and choose coordinates such that the z-axis coincides with the direction of travel of the GW as described in~\ref{sec: DetectorTensor}. The redshift can be trivially generalized to the case of multiple sources or a background by integrating over all directions.
\begin{align}
	h(t,\vec{x}) =& \frac{1}{2}\int \tilde{h}_A(f) e^{2\pi\mathrm{i}f\left(t-\frac{\hat{\Omega}\cdot\vec{x}}{c}\right) + \mathrm{i}\varphi_A} + c.c.\, df\, e^A \notag \\
	=& \frac{1}{4\pi}\int \tilde{h}_A(\omega)
	e^{\mathrm{i}\omega\left(t-\frac{\hat{\Omega}\cdot\vec{x}}{c}\right) + \mathrm{i}\varphi_A} + c.c.\, d\omega\, e^A \notag \\
	=& \frac{1}{2\pi}\int \tilde{h}_A(\omega) \cos\left(\omega\left[t-\frac{\hat{\Omega}\cdot\vec{x}}{c}\right]
	+ \varphi_A\right) d\omega\, e^A,
\end{align}
where $\tilde{h}_A\in C^\infty(\mathbb{R})$.\\

Since our derivation is not dependent on $\omega$ (we took derivatives and integrated with respect to $\lambda$ and $\tau$) we can generalize our result by simply replacing:
\begin{equation}
h_A \mapsto \frac{1}{2\pi}\int \tilde{h}_A(\omega) d\omega.
\end{equation}
So the $\Delta h_A^c(\tau_1)$-term becomes:
\begin{widetext}
	\begin{align}
		h_A\left(\cos(\varphi+\varphi_A)-\cos\Phi_A(\tau_1)\vphantom{\sqrt{2}}\right)
			\mapsto \frac{1}{2\pi}\int \tilde{h}_A(\omega) \left(\cos(\varphi+\varphi_A)-\cos\Phi_A(\tau_1)\vphantom{\sqrt{2}}\right) d\omega.
	\end{align}
\end{widetext}
using $\varphi = \omega t$ we can identify
\begin{align}
	\Delta h_A^c(\tau_1) =& h_A^c(\bold{x}^\mu(\lambda_E)) - h_A^c(\bold{x}^\mu(\lambda_a)) \\
	=& h_A^c\left(t+\mathcal{O}(h),\vec{x}_E\right) - h_A^c\left(t-\frac{L}{c}+\mathcal{O}(h),\vec{x}_a\right) \notag \\
	\overset{\mathcal{O}(h)}{\approx}& h_A(t) - h_A\left(t - \frac{L}{c}[1+\gamma]\right) = h_A(t) - h_A(t_a), \notag
\end{align}
where $t$ is the photons arrival time at Earth and $t_a$ when it was emitted from the pulsar.\\

In the case of the sine terms $\Delta h_A^s(\tau_1)$ the situation is a bit less obvious:
\begin{widetext}
	\begin{equation}
		h_A\left(\sin(\varphi+\varphi_A)-\sin\Phi_A(\tau_1)\vphantom{\sqrt{2}}\right)
			\mapsto \frac{1}{2\pi}\int \tilde{h}_A(\omega) \left(\sin(\varphi+\varphi_A)-\sin\Phi_A(\tau_1)\vphantom{\sqrt{2}}\right) d\omega.
	\end{equation}
\end{widetext}
We could use $\sin(x) = \cos(x-\frac{\pi}{2})$ to compare to the original wave, however then we would have to make a substitution which will change $\tilde{h}_A$:\\
\begin{align}
\int \tilde{h}_A(\omega) \cos\left(\omega t - \frac{\pi}{2}\right) d\omega = \int \tilde{h}_A\left(\frac{\eta+\frac{\pi}{2}}{t}\right)
	\cos\eta \frac{d\eta}{t}.
\end{align}
Instead we take the derivative after $t$:
\begin{widetext}
	\begin{align}
		\Delta h_A^s(\tau_1) = -d_t\int \tilde{h}_A(\omega) \left(\cos(k_\mu x_E^\mu + \varphi_A)
			- \cos(k_\mu x_a^\mu + \varphi_A)\vphantom{\sqrt{2}}\right) d\omega = -\left[\dot{h}_A(t)-\dot{h}_A(t_a)\right],
	\end{align}
\end{widetext}
where the dot denotes derivative after coordinate time $t$ here and not after the parameter $\lambda$ as above.\\

Finally, the full redshift formula to first order in $h$ is given by:
\begin{widetext}
	\begin{align}
		z_P(t) \approx& F^A \frac{\Delta \dot{h}_A(t) - \Delta \dot{h}_A(t+\omega T_a)}{\omega T_a[1+\gamma]} \\
		 &- \frac{\alpha_\omega^k \varepsilon_{kij}}{2\pi}
			\frac{\alpha^i-\alpha_\omega^i (\vec{\alpha}\cdot\vec{\alpha}_\omega)}
				{\vert \vec{\alpha} - \vec{\alpha}_\omega (\vec{\alpha}\cdot\vec{\alpha}_\omega) \vert^2}
			\left\lbrace c\frac{I^j F^{jA} - \alpha_\omega^j \alpha_{\omega l} I^l F^{lA}}{L\omega[1+\gamma]^2}
			\left( \Delta \dot{h}_A(t) - \Delta \dot{h}_A(t+\omega T_a) \vphantom{\sqrt{2}}\right) \right. \notag \\
		 &\left. + \left(\frac{1}{2}\left(\mathcal{F}^{jA} + \alpha_\omega^j \alpha_{\omega l} \mathcal{F}^{lA}\right)
			- \frac{I^j F^{jA} - \alpha_\omega^j\alpha_{\omega l} I^l F^{lA}}{1+\gamma}\right)
			\left(h_A(t_a + \omega T_a) - h_A(t_a)\vphantom{\sqrt{2}}\right) \right\rbrace + \mathcal{O}(h^2),
	\end{align}
	with
	\begin{equation}\label{eq.: Delta h}
		\Delta h_A(t) = h_A(t) - h_A(t_a), \quad t_a = t - \frac{L}{c}[1+\gamma].
	\end{equation}

	And the expansion to first order in $\omega T_a$ as well is:
	\begin{align}
		z_P(t) \approx& \frac{F^A \Delta h_A(t)}{1+\gamma}
			+ \frac{\omega T_a}{2} \frac{F^A \Delta \dot{h}_A(t)}{1+\gamma} - \frac{\omega T_a}{2\pi} \alpha_\omega^k \varepsilon_{kij}
			 \frac{\alpha^i - \alpha_\omega^i (\vec{\alpha}\cdot\vec{\alpha}_\omega)}
				{\vert \vec{\alpha} - \vec{\alpha}_\omega (\vec{\alpha}\cdot\vec{\alpha}_\omega) \vert^2}
			\left\lbrace c\frac{I^j F^{jA} - \alpha_\omega^j \alpha_{\omega l} I^l F^{lA}}{L\omega[1+\gamma]^2}
			\Delta h_A(t) \right. \notag \\
		 &\left. - \left(\frac{1}{2} \left(\mathcal{F}^{jA} + \alpha_\omega^j\alpha_{\omega l} \mathcal{F}^{lA}\right)
			 -\frac{I^j F^{jA} - \alpha_\omega^j \alpha_{\omega l} I^l F^{lA}}{1+\gamma}\right) \dot{h}_A(t_a) \right\rbrace + \mathcal{O}(h^2) + \mathcal{O}(h(\omega T_a)^2).
	\end{align}
\end{widetext}

We observe that the first correction term in $h\,\omega T_a$ is as usual proportional to the pattern functions. The second one however only contains the generalized versions since this term does not come from the time component of the photon geodesic but from its spatial components and the spatial direction of the rotation axis of the pulsar. It has a different structure, than the previous terms, since it describes a delay/speedup coming from the fact that the pulsar does not precisely make one rotation until the next photon, which will hit Earth, is emitted and thus is not related to the travel time of the photon.

\section{Discussions}\label{sec. discussions}
After pointing out, that there is a logical difference between the redshift of the frequency of a single photon and the redshift of the pulses, which are subsequent streams of photons, we derive the pulse redshift to first order in the strain amplitude, under the influence of a generic spin 2 field. We make no prior assumptions about the nonexistence of cross terms between different polarizations due to linearity and instead find it as a result of considering all six polarizations. Due to this generality our derivation can be straightforwardly extended to second order in the strain.\\

The higher order terms in $\omega T_a$ come from the fact that we are calculating the pulse redshift i.e. arrival time difference versus photon redshift, which is obtained from the difference of the time component of the photon momentum at Earth and pulsar. The last term, also proportional to $\omega T_a$, comes from the fact that the pulsar does not exactly make a full rotation until the next photon is released, which will arrive at Earth.\\
The IPTA collaboration~\cite{IPTA,NANOGRav11yr,EPTA,PPTA} measures in the frequency range between \SI{1}{nHz} and \SI{100}{nHz}: $f \in [10^{-9},10^{-7}]\SI{}{Hz} = [10^{-9},10^{-7}]\cdot10^{-3}\SI{}{ms^{-1}}$. They only use millisecond pulsars $T_a \approx \SI{1}{ms}$ and thus the GW angular frequency times pulsar period is in the range of $\omega T_a \approx f T_a \in [10^{-12},10^{-10}]$, ($2\pi \approx 1$). The strain of the GW signals for which PTA's aim for is expected to be of order $h \approx 10^{-15}$.\\
So, it makes sense to expand only to zeroth order in $h\omega T_a$. If one would however include slower rotating pulsars, there are some with a period up to \SI{23.5}{s}~\cite{SlowPulsar} ($\omega T_a\in [10^{-8},10^{-6}]$) and attempt to measure gravitational waves at higher frequencies, then the redshift formula without this expansion as given in~\eqref{eq.: z_P} can become instrumental. Currently every pulsar is measured roughly once a month, so one could in principle increase the frequency range by orders of magnitude. To do this one would require a higher cadence which would also improve the sensitivity. One could even go so far as to measure continuously, and with a long enough observation time one could even observe above the pulsar frequencies by matching the resulting pseudo random pulse redshift. Since the pulsars have only white noise and red noise (for some cases) there are no obvious physical limits other than the observation time and cadence limiting high frequency sensitivity.\\

The $(1+\gamma)$-term appears frequently in the denominator of the redshift formula. This has a pole at $\gamma \to -1$. These poles are always cancelled by $\Delta h_A$~\eqref{eq.: Delta h} in the numerator which goes to zero faster then $1+\gamma$ goes to infinity for $\gamma\to-1$. One can be tempted to look in the pattern functions for such a counter term but this only works for the $b$ polarization. However, for $+$ and $\times$ the counter term in the pattern function gives finite values but still leaves a discontinuity, whilst for the case of $x,y$ and $l$ polarizations the terms in the pattern function fail to cancel these poles. Thus we decide to not include the $(1+\gamma)$-term into the pattern functions. With this choice, our pattern functions agree with the definition used for interferometers in the case of a single arm detector. Instead it can be seen as the denominator of a term describing an interference of the gravitational wave with the photon geodesic and thus is dependent on the angle $\gamma = \hat{\Omega}\cdot\hat{p} = \cos\theta_\Omega$ between the travel direction of the GW and the one of the photon.\\
When one calculates the $SNR$ (signal-to-noise ratio) for a gravitational wave background, one collects all geometric terms, integrated over all directions, into a function called the overlap reduction function. As we have pointed out at the end of section~\ref{subsec: H&D} this interference provides geometry dependence via the $(1+\gamma)$-denominator. This causes a pole for each pulsar in the direction integral. The problematic points are the ones exactly behind the pulsars and since a background is a signal coming from all directions these two poles cannot simply be removed from the integral. Since the short wavelengths approximation cannot be applied around these poles we will calculate the overlap reduction function for the tensor mode without this approximation in our next paper.

\begin{acknowledgements}
We thanks the anonymous referee for many useful comments and suggestions. A.B. is supported by the Tomalla Foundation,  S.T. is supported by Swiss National Science Foundation grant number 200020 182047.\\
\end{acknowledgements}

\appendix
\section{The source aligned with the pulsar}
If the source is aligned with the pulsar $\gamma = -1$ which leads to poles where we divide by $[1+\gamma]$. We can however just solve the initial value problem again for this special case. The geodesic equations simplify to:
\begin{equation}
\delta\dot{p}^\mu(\lambda) \approx \frac{\mathcal{P}^2\omega}{c}I^\mu F^{\mu A}h_A\sin(\varphi+\varphi_A).
\end{equation}
Since the right hand side is constant in $\lambda$ we get a linear function for the momentum and a quadratic one for geodesic:
\begin{align}
	p^\mu(\lambda) =& p_0^\mu + \delta p_0^\mu + \frac{\mathcal{P}^2\omega}{c}I^\mu F^{\mu A}h_A\sin(\varphi+\varphi_A)\lambda, \notag \\
	x^\mu(\lambda) =& x_0^\mu + \delta x_0^\mu + (p_0^\mu + \delta p_0^\mu)\lambda \notag \\
	&+ \frac{\omega}{2c}\mathcal{P}^2\lambda^2I^\mu F^{\mu A}h_A\sin(\varphi+\varphi_A).
\end{align}
Fixing the boundary conditions in the same way at Earth and pulsar gives us $\delta x_0^i = 0$ and:
\begin{align}
	&x^i(\lambda_a) = L\alpha^i - \frac{L}{\mathcal{P}}\delta p_0^i
	+ \frac{L^2\omega}{2c}I^iF^{iA}h_A\sin(\varphi+\varphi_A) = L\alpha^i \notag \\
	&\Rightarrow \quad \delta p_0^i = \mathcal{P}\frac{L\omega}{2c}I^iF^{iA}h_A\sin(\varphi+\varphi_A).
\end{align}
Thus, the spatial part is given by:
\begin{align}
p^i(\lambda) &= -\mathcal{P}\left( \alpha^i 
	- \frac{\omega}{c}\left[\frac{L}{2}+\mathcal{P}\lambda\right]I^iF^{iA}h_A\sin(\varphi+\varphi_A) \right), \notag \\
x^i(\lambda) &= -\mathcal{P}\left( \alpha^i\lambda
	- \frac{\omega}{2c}\left[L\lambda+\mathcal{P}\lambda^2\right]I^iF^{iA}h_A\sin(\varphi+\varphi_A) \right).
\end{align}
The null-line condition yields:
\begin{equation}
\delta p_0^t=\frac{P}{c}\left(F^{tA} h_A\cos(\varphi+\varphi_A) + \frac{L\omega}{2 c}F^{tA}h_A\sin(\varphi+\varphi_A)\right),
\end{equation}
and by fixing $\delta t_0$ in the same way as above we get:
\begin{equation}
\delta t_0 = \frac{L}{c}F^{tA} h_A \cos(\varphi+\varphi_A),
\end{equation}
which determines the time-like part:
\begin{align}
	t(\lambda) =& \frac{\varphi}{\omega} + [L+P\lambda]\frac{F^{tA}}{c} h_A \cos\left(\varphi+\varphi_A\right) \notag \\
	&+ P\left(\frac{\lambda}{c}+\frac{\omega}{2 c^2}\left[L\lambda+P\lambda^2\right] F^{tA} h_A \sin\left(\varphi+\varphi_A\right)\right),
	\notag \\
	p^t(\lambda) =& \frac{P}{c}\left(1 + F^{tA} h_A\cos\left(\varphi+\varphi_A\right) 
		\vphantom{\frac{L}{c}}\right.\notag \\
	&\left. + \frac{\omega}{c}\left[\frac{L}{2}+P\lambda\right] F^{tA} h_A\sin\left(\varphi+\varphi_A\right)\right).
\end{align}

We can check that this coincides with the limit of the general case. For $\gamma\to-1$ we have $J^\mu = I^\mu$ and due to cancellations, the expression splits into different cases:
\begin{align}
	\lim_{\gamma\to-1}t(\lambda) =& \frac{\varphi}{\omega} + \frac{\mathcal{P}\lambda}{c}
		+ [L+\mathcal{P}\lambda]I^tF^{tA}\cos(\varphi+\varphi_A) \notag \\
	&+ \frac{\mathcal{P}\omega}{2c}\left[L\lambda+\mathcal{P}\lambda^2\right]I^tF^{tA}h_A\sin(\varphi+\varphi_A),
		\notag \\
	\lim_{\gamma\to-1}x(\lambda) =& \lim_{\gamma\to-1}y(\lambda) = 0, \quad \text{since} \quad \alpha = \beta = 0, \notag \\
	I^x =& I^y = -[1+\gamma] = 0, \quad \gamma = -1, \\
	\lim_{\gamma\to-1}z(\lambda) =& \mathcal{P}\left( \lambda + \frac{\omega}{2c}\left[L\lambda+\mathcal{P}
		\lambda^2\right] F^{zA}h_A\sin\varphi_A \right), \notag
\end{align}
for the geodesic and
\begin{align}
	\lim_{\gamma\to-1}p^t(\lambda) &= \frac{\mathcal{P}}{c}\left( 1 + F^{tA}\left\lbrace\vphantom{\frac{L}{c}}
		h_A\cos(\varphi+\varphi_A) \right.\right. \\
	&\left.\left. + \frac{\omega}{c}\left[\frac{L}{2}+P\lambda\right] h_A\sin(\varphi+\varphi_A)
		\right\rbrace \right), \notag \\
	\lim_{\gamma\to-1} p^x(\lambda) &= \lim_{\gamma\to-1} p^y(\lambda) = 0, \notag \\
	\lim_{\gamma\to-1} p^z(\lambda) &= \mathcal{P}\left( 1 + \frac{\omega}{c}\left[ \frac{L}{2} 
		+ \mathcal{P}\lambda \right]F^{zA}h_A\sin(\varphi+\varphi_A) \right), \notag
\end{align}
for the momentum.\\

We observe that the momentum has a linear term in $\lambda$ which causes a quadratic one in the geodesic. So, the photon seems to be accelerated. But on the other hand, we imposed the null line condition on it so it must propagate with light speed. To investigate this further we calculate the velocity of the photon in our chosen reference frame (pulsar and Earth at rest).\\
We invert $t(\lambda)$ via perturbation ansatz $\lambda = \lambda^{(0)} + \delta\lambda$ to calculate the velocity in that frame, using chain rule:
\begin{align}
	\lim_{\gamma \to -1}\frac{dz}{dt} &= \lim_{\gamma \to -1}\frac{dz}{d\lambda}\frac{d\lambda}{dt} \notag \\
	&= c\left(1 - \frac{h_l}{\sqrt{2}}\cos\left(\varphi_l+\varphi_a\right)\right)  + \mathcal{O}\left(h^2\right).
\end{align}
We see that the photon depending on the phase of the GW is moving faster or slower than light as seen by Earth by an order of $h$.

\section{Curved cross product components}\label{Ap: curvedCross}
The components in local coordinates are defined as:
\begin{align}
	&\mathcal{E}_p(\partial_i(p),\partial_j(p)) = \mathcal{E}^k_{ij}(p)\partial_k(p), \notag \\
	&\partial_i\in Der_p(U)\simeq T_pU, \quad i,j,k\in\{1,2,3\},
\end{align}
where the spatial partial derivative form a basis of the spatial tangent space $T_pU$ at the event $p$.\\

We write the properties \eqref{eq: sinprop} and \eqref{eq: perp} in the form of \eqref{eq: cross}:
\begin{equation}\label{eq: curvedCross}
\mathcal{E}_p(X,Y) = \Vert X\Vert \Vert Y\Vert \sin\theta\, \hat{n},
\end{equation}
where $\theta$ is ensured to be consistent with equation \eqref{eq: skp} by property \eqref{eq: sinprop}.\\

To get the cross-product coefficients dependent on the metric we can use $\sin\theta=\sqrt{1-\cos^2\theta}$ and replace the cosine with equation \eqref{eq: skp}:
\begin{align}\label{eq: cC(g)}
	\mathcal{E}_p(X,Y) &= \Vert X\Vert \Vert Y\Vert \sqrt{1-\cos^2\theta}\, \hat{n} \notag \\
	&= \Vert X\Vert \Vert Y\Vert \sqrt{1-\frac{\bar{g}_p(X,Y)^2}{\Vert X\Vert^2 \Vert Y\Vert^2}}\,\hat{n}.
\end{align}
The normal vector $\hat{n}$ takes care of the sign, which we neglected while replacing $\sin\theta$.\\

Suppressing the dependence on $p$ we determine the components by evaluating \eqref{eq: cC(g)} on the basis:
\begin{align}
\mathcal{E}(\partial_i,\partial_j) =& \sqrt{\bar{g}(\partial_i,\partial_i)\bar{g}(\partial_j,\partial_j)
	- \bar{g}(\partial_i,\partial_j)^2} n_{ij}^k\partial_k \notag \\
\Rightarrow \quad \mathcal{E}^k_{ij} =& \sqrt{\bar{g}_{ii}\bar{g}_{jj}-\bar{g}_{ij}^2}n_{ij}^k,
\end{align}
where $n_{ij}^k\partial_k = \hat{n}(\partial_i,\partial_j)$ is the normal vector perpendicular to $\partial_i$ and $\partial_j$. Thus it must satisfy:
\begin{align}
&\bullet\quad \bar{g}(\partial_i,\hat{n}(\partial_i,\partial_j)) = n_{ij}^k\bar{g}_{ik} = 0 \tag{I} \label{eq: I}\\
&\bullet\quad \bar{g}(\partial_j,\hat{n}(\partial_i,\partial_j)) = n_{ij}^k\bar{g}_{jk} = 0 \tag{II} \label{eq: II}\\
&\bullet\quad \bar{g}(\hat{n}(\partial_i,\partial_j),\hat{n}(\partial_i,\partial_j)) = n_{ij}^kn_{ij}^l\bar{g}_kl = 1
	\tag{III} \label{eq: III}
\end{align}
We have three components and three equations where the last one is quadratic and thus $\hat{n}$ is defined up to a sign which is determined by property \eqref{eq: flatlim}.\\

Our strategy for solving these equations is to write the sum over $k$ explicitly (plug in numbers for $k$) and solve them in an inductive way, so we can read off the general structure of the solution.
\begin{align}
n_{ij}^1\bar{g}_{i1} &\overset{\eqref{eq: I}}{=} - n_{ij}^2\bar{g}_{i2} - n_{ij}^3\bar{g}_{i3},
	&\qquad n_{ij}^1\bar{g}_{j1} &\overset{\eqref{eq: II}}{=} - n_{ij}^2\bar{g}_{j2} - n_{ij}^3\bar{g}_{j3}, \notag \\
n_{ij}^2\bar{g}_{i2} &\overset{\eqref{eq: I}}{=} - n_{ij}^3\bar{g}_{i3} - n_{ij}^1\bar{g}_{i1},
	&\qquad n_{ij}^2\bar{g}_{j2} &\overset{\eqref{eq: II}}{=} - n_{ij}^3\bar{g}_{j3} - n_{ij}^1\bar{g}_{j1}, \notag \\
n_{ij}^3\bar{g}_{i3} &\overset{\eqref{eq: I}}{=} - n_{ij}^1\bar{g}_{i1} - n_{ij}^2\bar{g}_{i2},
	&\qquad n_{ij}^3\bar{g}_{j3} &\overset{\eqref{eq: II}}{=} - n_{ij}^1\bar{g}_{j1} - n_{ij}^2\bar{g}_{j2}. \notag
\end{align}

Now we can construct new equations, by adding and subtracting \eqref{eq: I} and \eqref{eq: II}:
\begin{align}
	n_{ij}^1(\bar{g}_{i1}+\bar{g}_{j1}) &= - n_{ij}^2(\bar{g}_{i2}+\bar{g}_{j2}) - n_{ij}^3(\bar{g}_{i3}+\bar{g}_{j3}),
		\notag \\
	n_{ij}^2(\bar{g}_{i2}+\bar{g}_{j2}) &= - n_{ij}^3(\bar{g}_{i3}+\bar{g}_{j3}) - n_{ij}^1(\bar{g}_{i1}+\bar{g}_{j1}),
		\notag \\
	n_{ij}^3(\bar{g}_{i3}+\bar{g}_{j3}) &= - n_{ij}^1(\bar{g}_{i1}+\bar{g}_{j1}) - n_{ij}^2(\bar{g}_{i2}+\bar{g}_{j2}),
		\notag
\end{align}
\begin{align}
	n_{ij}^1(\bar{g}_{i1}-\bar{g}_{j1}) &= n_{ij}^2(\bar{g}_{j2}-\bar{g}_{i2}) + n_{ij}^3(\bar{g}_{j3}-\bar{g}_{i3}), 	
		\notag \\
	n_{ij}^2(\bar{g}_{i2}-\bar{g}_{j2}) &= n_{ij}^3(\bar{g}_{j3}-\bar{g}_{i3}) + n_{ij}^1(\bar{g}_{j1}-\bar{g}_{i1}), 
		\notag \\
	n_{ij}^3(\bar{g}_{i3}-\bar{g}_{j3}) &= n_{ij}^1(\bar{g}_{j1}-\bar{g}_{i1}) + n_{ij}^2(\bar{g}_{j2}-\bar{g}_{i2}). 
		\notag
\end{align}

\begin{widetext}
	Comparing the added and subtracted equations we can eliminate one of the three variables:
	\begin{align}
		n_{ij}^1 &= -\frac{n_{ij}^2(\bar{g}_{i2}+\bar{g}_{j2}) + n_{ij}^3(\bar{g}_{i3}+\bar{g}_{j3})}{\bar{g}_{i1}+\bar{g}_{j1}}
			= \frac{n_{ij}^2(\bar{g}_{j2}-\bar{g}_{i2} + n_{ij}^3(\bar{g}_{j3}-\bar{g}_{i3}))}{\bar{g}_{i1}-\bar{g}_{j1}}, \notag \\
		n_{ij}^2 &= -\frac{n_{ij}^3(\bar{g}_{i3}+\bar{g}_{j3}) + n_{ij}^1(\bar{g}_{i1}+\bar{g}_{j1})}{\bar{g}_{i2}+\bar{g}_{21}}
			= \frac{n_{ij}^3(\bar{g}_{j3}-\bar{g}_{i3} + n_{ij}^1(\bar{g}_{j1}-\bar{g}_{i1}))}{\bar{g}_{i2}-\bar{g}_{j2}}, \notag \\
		n_{ij}^3 &= -\frac{n_{ij}^1(\bar{g}_{i1}+\bar{g}_{j1}) + n_{ij}^2(\bar{g}_{i2}+\bar{g}_{j2})}{\bar{g}_{i3}+\bar{g}_{j3}}
			= \frac{n_{ij}^1(\bar{g}_{j1}-\bar{g}_{i1} + n_{ij}^2(\bar{g}_{j2}-\bar{g}_{i2}))}{\bar{g}_{i3}-\bar{g}_{j3}}. \notag
	\end{align}
\end{widetext}

We can simplify these expressions to get a direct comparison between two components:
\begin{align}
	n_{ij}^2(\bar{g}_{i1}\bar{g}_{j2}-\bar{g}_{i2}\bar{g}_{j1}) = n_{ij}^3(\bar{g}_{i3}\bar{g}_{j1}-\bar{g}_{i1}\bar{g}_{j3}), \\
	n_{ij}^3(\bar{g}_{i2}\bar{g}_{j3}-\bar{g}_{i3}\bar{g}_{j2}) = n_{ij}^1(\bar{g}_{i1}\bar{g}_{j2}-\bar{g}_{i2}\bar{g}_{j1}), \\
	n_{ij}^1(\bar{g}_{i3}\bar{g}_{j1}-\bar{g}_{i1}\bar{g}_{j3}) = n_{ij}^2(\bar{g}_{i2}\bar{g}_{j3}-\bar{g}_{i3}\bar{g}_{j2}).
\end{align}\\

Now we use these identities to express equation \eqref{eq: III} in terms of a single component. We can then solve for this component and find the other components using the identities above again. If we pick the sign such that $n_{ij}^k$ reduces to $\epsilon^{kij}$ for $\bar{g}_{ij} = \delta_{ij}$, as required by \eqref{eq: flatlim}, we get:
\begin{widetext}
	\begin{align}
		n_{ij}^k &= \frac{1}{N_{ij}}\epsilon^{kmn}\bar{g}_{mi}\bar{g}_{nj}, \\
		N_{ij} &= \sqrt{\det\left(\bar{g}_{i:}\wedge\bar{g}_{j:}+\text{diag}(\bar{g})\vphantom{\sqrt{2}}\right)
			- \det\left(\text{diag}(\bar{g})\vphantom{\sqrt{2}}\right)
			+ \sum_{k<l}(-1)^{l-k}\det([\bar{g}_{i:}\wedge\bar{g}_{j:}]_{kl})}, \notag
	\end{align}
\end{widetext}
where $\bar{g}_{i:} = (\bar{g}_{ij})_{j=1}^3$ denotes the $i$-th row or column of $\bar{g}$, $\text{diag}(\bar{g}) = (\bar{g}_{mm}\delta_{mn})_{mn}$ the matrix containing only the diagonal elements and $[\bar{g}_{i:}\wedge\bar{g}_{j:}]_{kl}$ is the matrix $\bar{g}_{i:}\wedge\bar{g}_{j:} = \bar{g}_{i:}\otimes\bar{g}_{j:} - \bar{g}_{j:}\otimes\bar{g}_{i:}$ where the $(k,l)$- and the $(l,k)$-components are replaced with $\bar{g}_{kl}$:\\
\begin{equation}
([\bar{g}_{i:}\wedge\bar{g}_{j:}]_{kl})_{mn} \coloneq 
\begin{cases}
	(\bar{g}_{i:}\wedge\bar{g}_{j:})_{mn}, & \{m,n\} \neq \{k,l\} \\
	\bar{g}_{kl}, & \{m,n\} = \{k,l\}
\end{cases}
\end{equation}

One can also express $N_{ij}$ as norm of the direction vector to which the cross-product points:\\
\begin{equation}
N_{ij} = \Vert(\epsilon^{kmn}\bar{g}_{mi}\bar{g}_{nj})_k\Vert = \sqrt{\bar{g}_{kl}\epsilon^{kmn}\bar{g}_{mi}\bar{g}_{nj}\epsilon^{lab}\bar{g}_{ai}\bar{g}_{bj}}.
\end{equation}

Putting everything together we can finally write down the formula for the components of the cross-product in a curved space:
\begin{align}
	\mathcal{E}^k_{ij} &= \frac{1}{N_{ij}} \sqrt{\bar{g}_{ii}\bar{g}_{jj}-\bar{g}_{ij}^2}
		\epsilon^{kmn}\bar{g}_{mi}\bar{g}_{nj} \notag \\
	&= \sqrt{\bar{g}_{ii}\bar{g}_{jj}-\bar{g}_{ij}^2} \frac{\epsilon^{kmn}\bar{g}_{mi}\bar{g}_{nj}}{\Vert(\epsilon^{kmn}\bar{g}_{mi}\bar{g}_{nj})_k\Vert}.
\end{align}

\section{Calculating photon geodesics}\label{Ap: Pgeod}
Not every 1-parameter null-vector field has a geodesic as integral curve. An obvious and rather silly example is a spiral null-line in Minkowsky space-time:\\
\begin{align}
    x(\lambda) = \cos\lambda, \quad y(\lambda) = \sin\lambda, \quad z(\lambda) = 0.
\end{align}
then the mass-shell equation reads:
\begin{align}
    &\bold{p}^2(\lambda) = \eta_{\mu\nu}\dot{x}^\mu\dot{x}^\nu = -\dot{t}^2 + \sin^2\lambda + \cos^2\lambda = 0 \\
    &\Rightarrow \quad t = \pm\lambda + \text{const.}
\end{align}
Dispite the tangent vectors being a null-vector field, this curve is no solution to the geodesic equations:
\begin{align}
    \ddot{t} = 0, \quad \ddot{x} = -\cos\lambda \neq 0, \quad \ddot{y} = -\sin\lambda \neq 0, \quad \ddot{z} = 0,
\end{align}
with 0 we mean the real zero function.\\

Another example is the vector field used in Chamberlin \& Siemens \cite{Chamberlin&Siemens} to calculate the photon redshift which is then used as the pulse redshift, whithout mentioning that approximation.\\
We show on an example, that the 1-parameter vector field
\begin{equation}
    \sigma^a = \nu\begin{pmatrix} 1 \\ -\alpha \\ -\beta \\ -\gamma\left( 1-\frac{h_L}{2} \right) \end{pmatrix}
\end{equation}
in a Minkowskiy space-time perturbed by a longitudinal wave of the form:
\begin{equation}
    g_{\mu\nu} = \eta_{\mu\nu} + h(t(\lambda)-z(\lambda))\, e^l_{\mu\nu}
\end{equation}
is not consistent with being the tangent vector field of a geodesic in that geometry.\\
To show that this is in general not a photon momentum we pick the direction $\alpha=\beta=0$, $\gamma=1$ and the wave-form $h(t-z) = h\cos(t-z)$. Then we get the following geodesic equations:
\begin{align}
    &d_\lambda\sigma^t = (\sigma^z)^2\frac{h}{2}\sin(t-z), \quad d_\lambda\sigma^x = 0, \notag \\  &d_\lambda\sigma^y = 0, \quad d_\lambda\sigma^z = (2\sigma^t-\sigma^z)\sigma^z\frac{h}{2}\sin(t-z)
\end{align}
inserting $\sigma^a = \nu\left(1,0,0,-\left( 1-\frac{h_L}{2} \right)\right)$ and expanding to first order in $h$ leads to:
\begin{align}\label{eq: sin=0}
    \frac{h}{2}\sin(t-z) + \mathcal{O}(h^2) = 0 \quad \Rightarrow \quad \sin(t-z) = 0 + \mathcal{O}(h)
\end{align}
for the $t$-component and
\begin{align}
    -\frac{3}{2}h\underbrace{\sin(t-z)}_{=0+\mathcal{O}(h)} + \mathcal{O}(h^2) = -\frac{h}{2}\underbrace{\sin(t-z)}_{=0+\mathcal{O}(h)}(\dot{t}-\dot{z}) + \mathcal{O}(h^2)
\end{align}
for the $z$-component, which is trivially satisfied due to~\eqref{eq: sin=0}. Furthermore this equation leads us to:
\begin{equation}\label{eq: t-z=0}
    t - z = n\pi + \mathcal{O}(h), \quad n\in\mathbb{Z}
\end{equation}
One can already see, that something is awry. The $t$- and the $z$- coordinates of the photon must be shifted by very specific numbers. That is not very physical.\\

We now integrate the null-vector to calculate the curve in question and check whether it satisfies the geodesic equations:
\begin{align}
    t(\lambda) &= \int \sigma^t d\lambda = \nu\lambda + A + \mathcal{O}(h^2), \notag \\
    z(\lambda) &= -\nu\int 1 - \frac{h}{2}\cos(\underbrace{t-z}_{=n\pi}) d\lambda \notag \\
    &= -\nu\lambda\left(1\mp\frac{h}{2}\right) + B + \mathcal{O}(h^2),
\end{align}
with $-$ for $n$ odd and $+$ if $n$ is even due to~\eqref{eq: sin=0} demanding the phase of the GW to be constant along the curve i.e. "photon surfing on the wave".\\
To satisfy the geodesic equations~\eqref{eq: t-z=0} must hold:
\begin{align}
    t(\lambda) - z(\lambda) &= \nu\lambda\left(2\mp\frac{h}{2}\right) + A + B + \mathcal{O}(h^2) \notag \\
    &\neq n\pi + \mathcal{O}(h) \qquad \lightning
\end{align}
We can set $A=B$ and $\frac{h}{2}$ is consistent with $\mathcal{O}(h)$ but we cant match a linear function to a constant.\\

Therefore one must solve the geodesic equations first (all 4 of them and not just the time component) and then solve the mass-shell equation to get a null-geodesic.\\
The geodesic equations are 4 coupled ordinary second order differential equations. This means, that they have 8 integration constants. These can be fixed by giving 8 initial-/ end-conditions or conditions on the momentum. In our derivation in section~\ref{sec. P.Geod.} we give 3 spatial initial and 3 end conditions which determine the starting point $\vec{x}_0+\delta\vec{x}_0$ and the initial spacial momentum $\vec{p}_0+\delta\vec{p}_0$. Then we solve for the initial time component of the photon momentum $p_0^t+\delta p_0^t$ using the mass shell equation $\bold{p}^2(\lambda) = m^2$, for $m = 0$. We use the remaining integration constant (initial position in time) as time variable and leave it unspecified, since photons are emitted from the pulsar at any time.

\bibliographystyle{apsrev4-1}
\bibliography{PTARef}

\end{document}